\documentclass[aps,prb,nobibnotes,twocolumn,nopacs,amsmath,amssymb]{revtex4-1}

\usepackage{amsmath}

\usepackage{graphicx}
\usepackage{dcolumn}
\usepackage{bm}
\usepackage{color}

\usepackage{amsmath}
\usepackage{amssymb}

\def\be{\begin{equation}}
\def\en{\end{equation}}
\def\bea{\begin{eqnarray}}
\def\ena{\end{eqnarray}}
\def\bec{\begin{equation}\begin{array}{rcl}}
\def\p{\partial}

\def\gs{\gtrsim}
\def\ls{\lesssim}

\def\ab{{ij}}

\newcommand{\av}[1]{\langle{#1}\rangle}

\newcommand{\bi}[1]{\mbox{\boldmath$#1$}}

\newcommand{\ppp}[3]{{\bigg(}\frac{\partial {#1}}{\partial {#2}}{\bigg )}_{#3}}

\begin{document}
\title{Theory of  electrolytes including  steric, 
attractive,  and hydration interactions  }
\author{Ryuichi Okamoto$^a$}
\email{okamoto-ryuichi@okayama-u.ac.jp}
\author{Kenichiro Koga$^{a,b}$}
\author{Akira Onuki$^c$}

\affiliation{
$^a$ Research Institute for Interdisciplinary Science, Okayama University, Okayama 700-8530, Japan \\
$^b$ Department of Chemistry, Faculty of Science, Okayama University, Okayama 700-8530, Japan \\
$^c$ Department of Physics, Kyoto University, Kyoto 606-8502, Japan 
}


\date{\today}

\begin{abstract} 
We present a continuum  theory of electrolytes 
composed of a waterlike solvent and univalent ions.  
First, we start with a density functional $\cal F$ 
 for the coarse-grained  solvent, cation, and anion densities,  
 including  the  Debye-H\"uckel free energy, the Coulombic interaction,  
and  the direct interactions among  these three  components. 
These densities  fluctuate  
obeying the distribution $\propto \exp(- {\cal F}/k_BT)$. 
Eliminating  the solvent density deviation in $\cal F$, 
 we obtain the    effective non-Coulombic interactions 
among the ions, which  consist  of the direct ones and  the 
solvent-mediated ones. 
We then  derive  general expressions for the ion correlation, 
the apparent partial  volume, and the activity  and  osmotic coefficients  
up to  linear order in the average salt density $n_{\rm s}$.  
Secondly, we perform numerical analysis  using   
 the  Mansoori-Carnahan-Starling-Leland model 
$[$J. Chem. Phys. {\bf 54}, 1523 (1971)$]$ for three-component hardspheres.  
The  effective   interactions sensitively depend 
on the cation and anion sizes due to competition between  
the steric and hydration effects,  
which are repulsive between  small-large ion pairs   
and attractive between symmetric pairs. 
These agree with previous experiments and   
 Collins' rule $[$Biophys. J. {\bf 72}, 65 (1997)$]$. 
We also give  simple approximate  expressions 
for the  ionic interaction coefficients valid for any ion sizes.  
\end{abstract}


\maketitle


\section{Introduction} 

The nature of how ions interact among
themselves and with water   has  been studied 
extensively in  physical chemistry\cite{RBook,Hamann}. 
In their seminal work in 1923,  Debye and H\"uckel\cite{Debye} (DH) 
 calculated the free energy correction  due to  the long-range 
 ion-ion correlation\cite{RBook,McQbook}.
To leading-order in the  average salt density $n_{\rm s}$, 
it is of order $n_{\rm s}^{3/2}$  and is 
determined by  the solvent dielectric constant $\epsilon$ 
and the ion valences, so it is exceptionally  {\it ion-nonspecific}. 
On the other hand, diverse phenomena sensitively 
 depend on the ion species in liquid water and 
 aqueous mixtures\cite{Kunz,Kunz2,NinhamReview}, 
where the short-range ion-ion and ion-solvent 
interactions come into play. 
Such {\it ion-specificity}   was originally reported  by  
Hofmeister\cite{[{}][{.~ English translation of 
Franz Hofmeister's historical papers
.}]Hof}  130 years ago in the 
 salting-out/salting-in effect of proteins. 
The  extended  DH theory\cite{Debye,Huckel,RBook,McQbook,Hamann} 
and  the Born theory of hydration\cite{Born,Millero,Marcus2011} 
already assumed certain ionic  radii specifically 
depending on the ion species. 

Since the early period of 
research\cite{Lewis,Bronsted,RBook,Hamann,Gu1,Gu2,Bromley,Pitzer,Ninham1}, 
there have been a great number of  measurements of  
the mean activity and  osmotic coefficients, 
$\gamma_{\pm}$ and $\varphi$. 
They have been  expanded  as 
 $1+ A\sqrt{n_{\rm s}}+ Bn_{\rm s}+\cdots$ for 
small $ n_{\rm s}$, where    the second term   
represents  the   DH part 
with an ion-nonspecific coefficient $A$. However,  
   the third term depends on the short-range interactions.  
and the coefficient    $B$  has been  
determined  empirically  for each ion pair. 
On the other hand,   the apparent partial volume  of 
salts\cite{RedlichJPC,RedlichReview,Millero,Marcus2011,DesnoyersT,Desnoyers}, 
written as $v_{\rm s}^{\rm ap}$,  
exhibits unique  ion-size-dependence  different   from  those of 
 $\gamma_{\pm}$ and $\varphi$.

In early {\it primitive} theories
\cite{Friedman,Lebowitz,Blum1,Blum2,McQbook,Ebeling,Fisher,Stell}, 
 the ions are hardspheres with charges $\pm q$, while  
the solvent is treated as a uniform continuum 
without any degrees of freedom (which much simplifies the 
calculations). 
Some simulations treated  cations 
and anions without solvent particles 
to confirm these theories\cite{Card,Pana}. 
We  also mention  general 
statistical mechanical studies \cite{Ramanathan,Blum1974,Pettitt,Dill,Case} 
and molecular dynamics (MD) 
simulations \cite{Smith1,VegtJCP, VegtPRL,Kalcher1, Kalcher2, Knecht,Netz1,Hasse,Naleem-Smith}, which attempted to take 
  into account the solvent effects in various manners.
Some simulations\cite{Smith1, Knecht,Netz1, Naleem-Smith}
 aimed to determine  the force-field parameters in simulation 
for each ion pair using the Kirkwood-Buff (KB) integrals\cite{Kirkwood}.
 From our viewpoint, it is still difficult to catch the overall 
physical picture of   the observed ion-specificity  
 from these papers.

As a key to the problem, 
 Widom {\it et al.}\cite{Widom1,Koga,Widom2} 
calculated  the second osmotic  virial coefficient $B_2= -G^0_{22}/2$ 
for a nonionic   solute in a one-component solvent\cite{McMillan}, 
where   $G^0_{22}$ is  the dilute limit of the 
solute-solute  KB  integral. 
Including the solvent degrees of freedom, they found    
\be 
B_2= B_2'' - (v_2^0- k_BT \kappa_{\rm w})^2 /{2k_BT \kappa_{\rm w}}, 
\en
where  $B_2''$ arises from  the direct 
 solute-solute interaction at a fixed solvent density. The  second volume term 
is due to the solvent-mediated interaction, 
where $v_2^0$ is the solute partial volume and  $ \kappa_{\rm w}$ is  
the solvent isothermal compressibility. It is largely negative  
 for nearly incompressible solvents with small $\kappa_{\rm w}$, leading to 
solute-solute attraction (particularly for large $v_2^0$). For electrolytes, 
the corresponding contributions have been missing in the previous  
theories\cite{Friedman,Lebowitz,Blum2,Blum1,Card,Ebeling,Fisher,Stell}.
In this paper, we extend  Eq.(1) to dilute electrolytes.

On electrolytes, there have been  numerous continuum  theories 
based on the  Poisson-Boltzmann equation in various situations
\cite{Evans1,Onuki,Bazant,Andel5,Fogolari}.  
To account for the excluded volumes, 
 the space-filling relation $\sum_i v_i n_i=1$ has been widely 
assumed\cite{Bik,Andelman,Ig,Onukibook},  
where $v_i$ is a  molecular volume  
of the $i$-th component with density $n_i$. 
Furthermore,  convenient is  a   
 continuum model of hardsphere mixtures by
Mansoori, Carnahan, Starling, and Leland (MCSL)\cite{MCSL}, 
as used in subsequent papers\cite{Bie,Bazant,Wa}. It 
is a generalization of the Carnahan and Starling 
 model of monodisperse  hardspheres\cite{Carnahan}.
Using  the MCSL model for neutral fluids, 
 we  studied 
small bubbles in water due to dissolved  
gases\cite{OkamotoOnukiBubble,OkamotoOnukiDFT} 
and phase behavior in ternary mixtures\cite{OkamotoOnukiTernary}
 such as  water-alcohol-hydrophobic solute\cite{Zemb}. 
 In the latter, the second term in Eq.(1) and another contribution from the concentration 
fluctuations were crucial.

In this paper, we first present a  statistical-mechanical 
theory setting  up    a free energy functional 
for the densities $n_1, n_2$, and $ n_3$ of 
the solvent, the cations, and the anions, respectively. 
Expressing the deviation $\delta n_1= n_1-\av{n_1}$ 
in terms of $n_2$ and $n_3$,  
 we   obtain the effective ion-ion interaction coefficients, 
written as $U_{ij}^{\rm eff}$ ($i, j=2,3$), which 
 have  bilinear volume terms as $B_2$ in Eq.(1). 
Using the continuum  MCSL and Born models, we show that 
 $U_{ij}^{\rm eff}$  
  tend to be  negative (attractive)  for symmetric ion pairs,  
 but tend to be positive (repulsive)  for small-large  pairs. 
These agree  with  experiments  
and  Collins' empirical rule\cite{Collins,Collins1,Collins2}.
Mathematically, the  total 
packing fraction  arises mainly from the solvent particles in our theory but  
 from the ions  only in the primitive  
theories\cite{Friedman,Lebowitz,Blum1,Blum2,McQbook,Ebeling,Fisher,Stell}. 
This leads to  largely  different results in the two approaches.

Small-large ion pairs exhibit unique behavior in water, 
which include   
 NaI as  a relatively mild example  and     NaBPh$_4$ as an extreme one. 
In the latter, tetraphenylborate   BPh$_4^-$ 
consists    of four phenyl rings bonded to an ionized 
boron\cite{Wipff,Taylor,MilleroTetra}. 
 In aqueous mixtures, adding a small amount of NaBPh$_4$ 
 is known to produce  mesophases due to preferential 
solvation\cite{Sadakane,Onuki2,Onuki,Yabunaka,RoijA}.

 The organization of this paper is as follows. In Sec.II, 
 we will start with a  free energy functional 
including  the  DH free energy. 
We will  then study the thermal density fluctuations 
accounting for  the solvent-mediated  correlations. 
In Sec.III, we will study the thermodynamics of electrolytes. 
  In Sec.IV,  we will first 
examine the ion volume and  the ion-ion interaction 
and then present  numerical analysis 
of various physical quantities.

\section{Fluctuations in electrolytes}

In our theory, the solvent is a  nearly incompressible, 
 one-component liquid, which is  also  called water, 
and the  ions have  the unit charges $\pm e$. 
 The salt or base added is assumed to 
 dissociate completely.  We do not treat  Bjerrum  
dipoles\cite{RBook,Bj,DS,Ebeling,Fisher,De,MaPair,Has,Fennell,Roij,Vegt3,A-B}
 as an independent  entity (see Appendix A). 
The effective ionic  diameters  are  not much 
 lager than that of the solvent  $d_1 (\cong 3 {\rm\AA}$ for water).  
 We  study  the  bulk  properties   without applied electric field. 
Thus, under   the periodic boundary condition, 
  the electrolyte is in a large $L\times L\times L$ box with 
   volume $V=L^3$. Generalization to the case of multivalent ions 
is straightforward\cite{McQbook} (see below Eq.(37)).
In this paper,  the temperature $T$ is fixed 
and its dependence of the physical quantities is  not  written explicitly. 

\subsection{Free energy functional $\cal F$  of  electrolytes}

We write  the coarse-grained number densities of water, cations, 
and anions as  $ n_1$, $ n_2$, and $ n_3$, respectively. Their  
  Fourier components  $n_i({\bi q})= \int d{\bi r}
n_i({\bi r})\exp(-{i{\bi q}\cdot{\bi r}})$  have  wave numbers smaller 
than an  upper  cut-off  $\Lambda$.
 In this section, assuming that  $\Lambda$ is smaller than 
the Debye wave number $\kappa$, 
we examine the thermal fluctuations of $n_i({\bi q})$ 
with $q<\Lambda$. They  obey  
the distribution $\propto \exp[-{\cal F}/k_BT]$, 
where we introduce    the     free energy  functional, 
\begin{align}
{\cal F} (\Lambda) 
=\int d{\bm r}  f + 
\frac{1}{2} \int d{\bi r}\rho \Phi.  
\end{align}
Here,  $f$ depends on $n_1$, $n_2$, and $n_3$ 
in the local density approximation.
The second term represents  the long-range Colombic 
intercation, where $\rho= e(n_1-n_2)$ is the charge density 
and $\Phi$ is  the electric potential  related  by       
$
-\nabla \cdot \epsilon{\nabla \Phi} = 4\pi\rho,
$ 
where $\epsilon$ is the dielectric constant.

We  expand  $ f$ up to the second order in  $n_2$ and $n_3$  as 
\bea 
&&\hspace{-5mm} 
 f  =f_{\rm w}( n_{1} ) +  
k_B T \sum_{{i}= 2,3} [\ln ( n_i\lambda_i^3) -1 + \nu_i(n_1)] n_i  
\nonumber\\
&& -\frac{1}{12\pi}  k_BT \kappa^3 + 
\frac{1}{2} \sum_{i,j= 2, 3} U_{\ab}( n_1) n_i n_j . \label{f_dilute}
\ena 
The first term  $ f_{\rm w}(n_{1} )$ is 
 the  free energy density of pure solvent. In the  second term,  
  $\lambda_i$ is the thermal de Broglie length 
and   $k_BT \nu_i(n_1)$    
is the  solvation chemical potential per ion due to 
 the  interactions between  an isolated 
ion of species $i$ and the solvent. The third term 
is the DH free energy density 
in the limit of low ion densities  
\cite{Debye,Huckel,McQbook,RBook}, where 
  $\kappa$ is the the  Debye wave number, 
\be 
\kappa = 
  [4\pi e^2(n_2+ n_3)/\epsilon(n_1) k_BT]^{1/2}. 
\en  
In the last term, $U_{ij}(n_1)$ 
represents the short-range  direct 
interactions between ion species $i$ and $j$ 
under influence of the solvent.  
Here,  $\epsilon(n_1)$,  $\nu_i(n_1)$,  and $U_{ij}(n_1) $ strongly depend 
on   $n_1$  in liquids.

 The DH free energy  can be  calculated  from the average of an 
excess  electric field around 
each ion, which is  produced by the other ions 
with separation distances shorter than $\kappa^{-1}$.  
Thus,  to use the DH theory,  we need to 
assume  $\Lambda<\kappa$.  Debye and H\"uckel  also introduced 
a  closest distance  around each ion in  the ion-ion 
correlation\cite{Debye,Huckel,RBook,McQbook,Hamann}, 
which  is written as  $a_2$ for the cations and as $a_3$ 
for the anions. The DH free energy density  is thus given by\footnote{
in the original paper\cite{Debye}, $a_2$ and $a_3$ can be different, 
while they have been equated  in  most subsequent papers.}      
\bea  
&&\hspace{-7mm} { f}_{\rm DH} = 
-\frac{1}{3} k_BT \ell_B\kappa \sum_{i=2,3}  n_i
 \tau(a_i \kappa)    \nonumber\\
&& = -\frac{1}{12\pi}k_BT \kappa^3 +  \frac{1}{2}  
\sum_{i,j=2,3}u_{ij}^{\rm ex}  n_i n_j +\cdots ,  
\ena 
where  $\tau(x) =3[\ln (1+x) -x+ x^2/2]/x^3$ and  
  and  $\ell_B=e^2/\epsilon(n_1)  k_BT$ is  
 the Bjerrum length ($= 7~{\rm \AA}$ in ambient water). 
In the second line,  
using  $\tau(x)=1- 3x/4+\cdots$ 
for $x\ll 1$, we write 
the  first correction for $a_i\kappa\ll 1$ with 
\be 
u_{ij}^{\rm ex}=  \pi k_BT \ell_B^2 (a_i + a_j) . 
\en  
Here, $u_{ij}^{\rm ex} = 34 k_BTd_1^3$ for 
$a_2=a_3=d_1=3~{\rm \AA}$ in ambient water. 
We assume that  $u_{ij}^{\rm ex}$ are included 
in   $U_{ij}$ in Eq.(3). In   Sec.IV, we will calculate  
the excess parts $ U_{ij}- u_{ij}^{\rm ex}$.

We suppose an 
equilibrium reference state, where the  average water and salt densities are 
written as  
\be 
\av{n_1}= n_{\rm w}, \quad 
\av{n_2}=\av{n_3}=n_{\rm s}.  
\en 
Under the overall charge neutrality,   we  
use  the mean solvation and interaction coefficients,  
\bea  
\nu &=& (\nu_2+\nu_3)/2, \\
U &=& ( U_{22}+ U_{33})/2+ U_{23} .
\ena   
We  also introduce   the  incompressibility  parameter,   
\be 
\epsilon_{\rm in}=n_{\rm w} k_BT \kappa_{\rm w},
\en 
where $\kappa_{\rm w}=1/(n_{\rm w}^2  \p^2 f_{\rm w}/\p n_{\rm w}^2) $ 
is the solvent isothermal compressibility. 
Here, $\epsilon_{\rm in}\ll 1$  
 for nearly incompressible liquids. 
For ambient  liquid water ($T= 300$~K and $p= 1$~atm),  
we have $\kappa_{\rm w}\cong 4.5\times 10^{-4}/$MPa and 
$\epsilon_{\rm in}\cong  0.062$.

\subsection{Thermal fluctuations and ion volumes   }

We here examine the long-wavelength density fluctuations 
to derive ion volumes. To this end,  we 
    superimpose small   density 
deviations  $\delta n_i({\bi r})$ on the averages  as
\be 
n_1= n_{\rm w}+\delta n_1, \quad 
n_i = n_{\rm s}+\delta n_i~(i=2,3).
\en  
where $\delta n_i$  have  Fourier components $n_i({\bi q})$ 
 with $q<\Lambda$. 

The deviation  
 $\delta{\cal F}={\cal F}-F$ of the free energy functional  
starts from  second-order terms 
 as\cite{OkamotoOnukiTernary} 
\begin{align}
\delta {\cal F}=\frac{1}{2} \int_{\bi q} 
\Big[\sum_{ i,j=1,2, 3} f_{ij}  n_i({\bi q})  n_j({\bi q})^* 
+\frac{4\pi }{\epsilon q^2} |\rho_{\bi q}|^2 \Big], 
\end{align}
where $\int_{\bi q}= V^{-1}\sum_{\bi q}$ represents the summation 
over  the wave vector ${\bi q}$. The second derivatives of 
$f$ with respect to the densities at fixed $T$ are written as    
\be 
f_{ij}={  \p^2 f}/ {\p n_i \p n_j} , 
\en  
which are the values at $n_1= n_{\rm w}$ and $n_2=n_3=n_{\rm s}$.  
In  Eq.(12), the Coulombic  term    
arises   from  the second term in  Eq.(2)  
with   $\rho_{\bi q}= e[  n_2({\bi q})- n_3({\bi q})]$.  
Then,    Eq.(3) gives 
\bea 
&&\hspace{-7mm}  
f_{11}={1}/({n_{\rm w}^2 \kappa_{\rm w}}) + 2k_BT\nu'' n_{\rm s}, \\
&&\hspace{-7mm} 
f_{1i}= k_BT [\nu_i'+ ({3}\epsilon'/4\epsilon)  \ell_B\kappa] 
+ ( U_{i2}'+ U_{i3}') n_{\rm s}, \\
&&\hspace{-7mm}f_{ij}={{k_BT}}
({\delta_{ij}}- \ell_B\kappa/8) /n_{\rm s}  + U_{ij},
\ena  
where $i,j=2,3$. Here, 
$\nu_i'= \p \nu_i/\p n_1$, $\nu''= \p^2 \nu/\p n_1^2$, 
 $\epsilon'= \p \epsilon/\p n_1$, and 
$U_{ij}'= \p U_{ij}/\p n_1$ at $n_1= n_{\rm w}$ (see  the value of 
 $\nu''$ for  NaCl below Eq.(45)). 
Data of $\epsilon$ for ambient water indicate\cite{Archer,Sengers} 
\be 
n_{\rm w}\epsilon'/\epsilon = \kappa_{\rm w}^{-1}  
(\p \ln \epsilon/\p p)_T =  1.1.
\en

In  the brackets in Eq.(12), the solvent-ion  coupling  
arises from    $[f_{12} n_2({\bi q}) + f_{13} n_3({\bi q})]  n_1({\bi q})^*$. 
Thus,    we   introduce the   deviation of the particle volume 
fraction\cite{OkamotoOnukiTernary},    
\bea 
&&\delta\phi_v = [\delta n_1+ (f_{12}/f_{11}) 
 \delta n_2+(f_{13}/f_{11}) \delta n_3]/n_{\rm w}
\nonumber\\
&&\hspace{5mm}\cong 
  n_{\rm w}^{-1}\delta n_1+ v_2^* \delta n_2+v_3^*\delta n_3.  
\ena 
The first line of Eq.(18) 
 can be used  for general  $n_{\rm s}$. 
In the second line  $v_i^*$ are    ion   volumes at infinite dilution,   
\be 
v_i^*  =\lim_{n_{\rm s}\to 0} f_{1i}/f_{11}n_{\rm w}
=\epsilon_{\rm in}  \nu_i' 
\quad (i=2,3).
\en  
For nonionic mixtures, 
$v_i^*$ corresponds to 
  $v_2^0- k_BT \kappa_{\rm w}$ in Eq.(1)\cite{Widom1,Koga,Widom2} 
and to $v_3^{\rm in}$  in our recent 
paper\cite{OkamotoOnukiTernary}. See also Eq.(21) and
the subsequent sentences.  

We can  then  rewrite $\delta{\cal F}$ in Eq.(12)  as   
\be
\delta {\cal F}=\frac{1}{2}{n_{\rm w}^2}{f_{11}}\int d{\bi r} |\delta\phi_v|^2 
+ \delta{\cal F}_{\rm ion}.
\en 
where  ${n_{\rm w}^2}{f_{11}}\cong \kappa_{\rm w}^{-1}$. 
Here, the  first term represents  the 
{\it steric interaction},  
 which    suppresses  the thermal fluctuations of  
$\delta\phi_v$   for small $\kappa_{\rm w}$. 
Namely, $\delta n_1$ tends to decrease by
 $n_{\rm w}(v_2^* \delta n_2+v_3^*\delta n_3)$ 
 on the average at long wavelengths. 
This interaction  can be derived  
 for  any  multi-component fluids
\cite{[{}][{.~ In this book, discussions 
are given on the fluctuation variances in Sec.1.3 and on 
the steric interaction in  polymer solution 
in Sec.3.5.}]Onukibook},  
 where    $\delta\phi_v \to 0 $ as  $\kappa_{\rm w}\to 0$.

The  volume $v_i^*$  is  of order  $d_i^3$ for 
large $d_i(>d_1$) in terms of  the   hardsphere diameter 
$d_i$, while it can be negative for small $d_i(<d_1$) 
 such as   Li$^+$   due to the 
hydration (see 
Sec.IIIF)\cite{Hepler,Mukerjee,Padova,Millero,Marcus2011,Craig}. 
From  measurements with  the  overall charge neutrality, 
  we can determine only  the  sum, 
\be 
v_{\rm s}^{*}= v_2^*+ v_3^*= 2\epsilon_{\rm in}\nu', 
\en  
where  $ \nu'= \p \nu( n_{\rm w})/\p n_{\rm w}$. 
This $v_{\rm s}^{*}$ is is smaller than the  corresponding 
infinite-dilution   partial volume ${\bar v}_{{\rm s}}^0$ 
in  Eq.(47) by $2k_BT \kappa_{\rm w}$. 
From  experimental reports on ${\bar v}_{{\rm s}}^0$ 
in  ambient   water\cite{Millero,Craig,MilleroTetra},  
 $n_{\rm w} v_{\rm s}^*  $ is $-0.21$,   0.93, 2.0,    and   15 
 for  LiF,    NaCl, NaI,   and    NaBPh$_4$, respectively.   Then, 
 $2n_{\rm w} \nu'$ is  $-3.4$,  $15$, 32,  and 240, respectively, 
 for these salts.   
 The  $\nu(n_{\rm w})$ itself  appears in the Henry constant.  

For nonionic mixtures,  the coefficients $f_{ij}$ are  written 
in terms of thermodynamic derivatives 
(see Eq.(26) in  our recent  paper\cite{OkamotoOnukiTernary}). 
Generally, $f_{ij}$ can be expressed  as  
\be 
f_{ij}/k_BT= \delta_{ij}/\av{n_i}- 
\int\hspace{-1mm} d{\bi r}  c_{ij}^{0}(  r ) \quad (i,j=1,2,3).
\en  
In terms of  the direct correlation functions 
$c_{ij}(r)$, we have   $c_{1j}^0(r)=c_{1j}(r)$ ($i=1$) 
and $c_{ij}^0(r) =c_{ij}(r)+  (-1)^{i+j} 
\ell_B /r$ ($i, j=2,3$)\cite{Oc,Evans1,Attard,Evans2,Hansen,Onukibook}. 
If  $f_{ij}$ are  defined in this manner, Eq.(12) can be used for general 
  $n_{\rm s}$. 
In the simple case of  a nonionic solute in one-component solvent, 
we  notice 
 $2k_BT  B_2=(\p \mu_2^{\rm ex}/\p n_2)_{T, \mu_1}=
 U_{22}^{\rm eff}$ 
and $2k_BT B_2''= 
(\p \mu_2^{\rm ex}/\p n_2)_{T, n_1}= 
- k_BT \int\hspace{-0.5mm} d{\bi r}  c_{22}(  r )
$ in Eq.(1), where $\mu_2^{\rm ex}$ is the excess 
solute chemical potential\cite{Koga,Widom2,OkamotoOnukiTernary}.

\subsection{Solvent-mediated interaction and Collins' rule }

Next, we  derive the solvent-mediated ion-ion interaction 
in the long wavelength.  To this end, we express 
 the   ionic  term in Eq.(20)   as 
\bea
&&\hspace{-5mm} 
\delta {\cal F}_{\rm ion}= {k_BT}
\int d{\bi r}  \Big[ \frac{| \delta n_2|^2+ |\delta n_3|^2}{2n_{\rm s}}  
- \frac{\ell_B \kappa}{16n_{\rm s}} {|\delta n_{\rm e}|^2 }\Big]
\nonumber\\
&&+\frac{1}{2}  \int_{\bi q}\Big[ 
\sum_{ i,j=2, 3} U_{ij}^{\rm eff}  n_i({\bi q})  n_j({\bi q})^* 
+\frac{4\pi }{\epsilon q^2} |\rho_{\bi q}|^2  \Big]. 
\ena 
In the first term,  $\delta n_{\rm e}= \delta n_2+\delta n_3$ is the 
ion density deviation. In the second term,   
we  introduce  the   effective ionic interaction coefficients, 
\be 
U_{ij}^{\rm eff}
= U_{ij}- v_i^* v_j^*/\kappa_{\rm w}\quad (i,j=2,3),
\en 
where the first term represents the short-ranged direct interactions and 
the  second   term arises from  
the solvent-mediated  interactions 
in the long wavelength limit. The second term 
  corresponds to  the second term in Eq.(1). 
The  Coulombic term in Eq.(23)    suppresses    
  $\rho_{\bi q}$  at small $q$.  
Thus, in thermodynamic quantities, there appears    the mean effective 
interaction  coefficient,   
\be 
 U_{\rm eff}= \frac{1}{2}\sum_{i,j=2,3}U_{ij}^{\rm eff}
=  U - \frac{1}{2\kappa_{\rm m}} ({v_{\rm s}^{*})^2}.    
\en

The second volume  term in Eq.(24)  is 
 amplified by $\kappa_{\rm w}^{-1}=n_{\rm w}k_BT/\epsilon_{\rm in}$ and 
is  very  large for not very small $v_i^*v_j^*$. However, it   
 does not appear if the solvent is treated as a homogeneous 
continuum\cite{Friedman,Lebowitz,Blum2,Blum1,Ebeling,Fisher,Stell}.
Indeed, it is needed to explain 
  Collins' rule\cite{Collins,Collins1,Collins2}.   
Namely,    if   $v_i^*$ and $v_j^*$ have  the same  sign, 
it is   negative 
leading   to     {\it solbophobic  attraction}  between species 
$i$ and $j$. See (a) and (b) in  Fig.1.  As a result, 
  this mechanism  yields   hydrophobic  
assembly of large solute 
particles\cite{Chandler,OkamotoOnukiTernary,Koga,Widom2}. 
On the other hand, 
 for  small-large ion pairs with  $v_2^* v_3^*<0$, 
 $U_{23}^{\rm eff} $  is positive  
leading to  non-Coulombic cation-anion   repulsion, 
as  in  Fig.1(c). See  Sec.IIIE  and Sec.IV for more 
analysis on the basis of Eq.(24). 
Previously, some attempts 
were made to explain Collins'  
rule not  using Eq.(24)\cite{Kalcher1,Fennell,Lund,Ninham2}.

We can also derive the second  term in Eq.(24) 
 in  the mean spherical  approximation (MSA) 
in the presence of 
 the solvent degrees of freedom\cite{Blum1974,Hansen}.
We also note   that the   interaction  energy  
in the Flory-Huggins  theory of polymer solutions   
corresponds to $n_{\rm w}U_{22}^{\rm eff}$ 
in our notation\cite{Onukibook}.

\begin{figure}[tbp]
\begin{center}
\includegraphics[width=190pt]{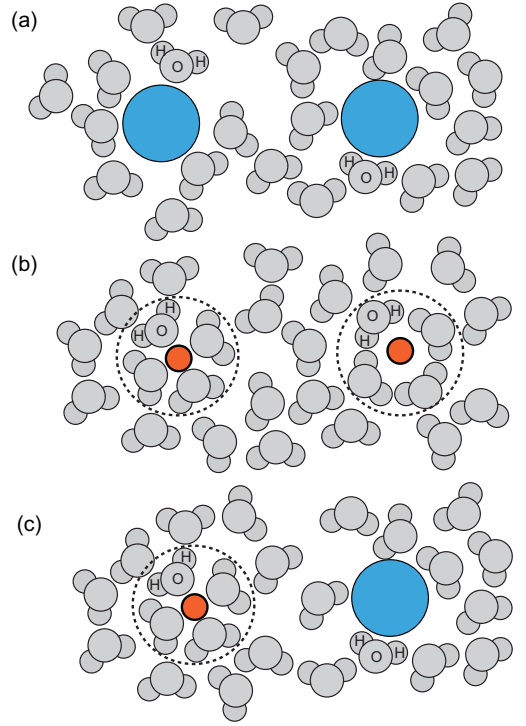}
\caption{(Color online) Illustration of two ions in close separation 
in water\cite{Collins}. (a) Large-large pair  with  non-Coulombic attraction. 
 Examples are CsI and CsBr. 
(b) Small-small   ions with  non-Coulombic attraction. 
 Examples are NaF and    LiF.  
(c)  Small-large (cation-anion) pair with  non-Coulombic repulsion.  
 Examples are NaI,   LiI, and NaBPh$_4$.  
Tendency of cation-anion association is 
 promoted with decreasing  $U_{23}^{\rm eff}$.  
 }
\end{center}
\end{figure}

\subsection{Fluctuation variances, charge density structure factor,  
 and  Kirwood-Buff integrals}

  We  treat $\delta n_i$ as the  thermal fluctuations   obeying  
the Gaussian distribution $\propto \exp(-\delta{\cal F}/k_BT)$. 
We  can then   calculate the  fluctuation variances 
 $I_{ij}= \lim_{q\to 0}
\av{n_{i}({\bi q})n_{j}({\bi q})^*}/V$, where 
  $L^{-1}\ll  q\ll \kappa$ in the limit of large $L$. 
Here, for any space-dependent
variables ${\hat A}({\bi r})$  and $\hat{B}({\bi r})$, we write 
$\av{\hat{A} : \hat{B}}= 
\lim_{q\to 0 }\av{A_{\bi q}{B_{\bi q}}^*}/V$, 
where $A_{\bi q}$ and $B_{\bi q}$ are the Fourier components\cite{Onukibook}. 
Then, $I_{ij}= \av{n_i:n_j}$. From  Eq.(20) we find\cite{OkamotoOnukiTernary}  
\bea
&&\av{\phi_v:\phi_v}=  k_BT/n_{\rm w}^2f_{11}\cong 
 \epsilon_{\rm in}/n_{\rm w},\\
&& \av{\phi_v : n_{i}}=0 ~~(i=2,3).  
\ena    
As  $q\to  0$, we have $\rho_{\bi q} \to  0$, so   we find   
\be 
 I_{22}=I_{33}=I_{23}= n_{\rm s} \chi,  
\en   
where     $ n_{\rm s}\chi$ represents  the amplitude 
 of the ion density fluctuations.  See its thermodynamic 
expression  in Eq.(44).  
From Eq.(18) we also find the solvent-solvent 
and solvent-ion fluctuation variances,  
\bea 
&&I_{11}=    k_BT/f_{11} + (f_{12}+f_{13})^2 f_{11}^{-2} 
n_{\rm s}\chi,\\
&&\hspace{-5mm} I_{12}=I_{13}=- (f_{12}+ f_{13})f_{11}^{-1} 
n_{\rm s}\chi \cong - v_{\rm s}^* n_{\rm w}n_{\rm s}/2 .  
\ena 
In  $I_{11}$, the first and second   terms are 
close to  $n_{\rm w}\epsilon_{\rm in}$ and 
$(n_{\rm w} v_{\rm s}^{*})^2   n_{\rm s}/2$, respectively, 
 for small $n_{\rm s}$.  Thus, the second one is dominant 
  for $n_{\rm s} /n_{\rm w}> 2\epsilon_{\rm in}/(v_{\rm s}^{*}n_{\rm w})^2$
($\sim 4\times 10^{-4}$ for NaPhB$_4$ in water), 
as    can be verified in   experiments.


It is convenient to  rewrite  
${\delta {\cal F}_{\rm ion}}$ in Eq.(23) in terms of 
 $\delta n_{\rm e}= \delta n_2+\delta n_3$ and $\rho= 
 e(\delta n_2-\delta n_3)$  as 
\bea
&&\hspace{-6mm} 
\frac{\delta {\cal F}_{\rm ion}}{k_BT}= 
\int d{\bi r}  \Big[ \frac{| \delta n_{\rm e}|^2}{8n_{\rm s}\chi}  
+(U_{22}^{\rm eff}-U_{33}^{\rm eff}) 
\frac{  {\delta n_{\rm e}}\rho}{4k_BTe} \Big]
\nonumber\\
&&\hspace{4mm} + \int_{\bi q}\Big[ 1+ w_{\rho}n_{\rm s} 
+\frac{\kappa^2 }{ q^2}\Big]\frac{  |\rho_{\bi q}|^2}{4e^2n_{\rm s}}. 
\ena 
The inverse  $\chi^{-1}$ in the first term 
depends on $n_{\rm s}$  as 
\be
\chi^{-1}= 2-  \ell_B \kappa/2 + 2n_{\rm s} U_{\rm eff} /k_BT .
\en 
The coefficient $w_\rho$ in the second term  arises from  asymmetry 
between the cations and the anions as   
\bea 
&&\hspace{-6mm} 
w_{\rho}=  
(U_{22}^{\rm eff}+U_{33}^{\rm eff}-2U_{23}^{\rm eff})/2k_BT \nonumber\\
&&\hspace{-4mm} 
= [U_{22}+ U_{33}-2U_{23}  - (v_2^*-v_3^*)^2
/2\kappa_{\rm w}]/2k_BT.
\ena

 From Eq.(31) the structure factor for the charge density 
fluctuations $\rho_{\bi q}$  for  $q \ll   \kappa$ 
 is given by\cite{Oc,Evans1,Evans2,Attard} 
\be 
S_{\rho\rho}(q) = 
\av{ |\rho_{\bi q}|^2}/V= 
{2e^2n_{\rm s}}  /[ 1+ w_{\rho}n_{\rm s}+{\kappa^2 }/{ q^2} ]. 
\en
 The cross term $\propto \rho\delta n_{\rm e}$  
in Eq.(31) gives a higher-order term$(\propto  n_{\rm s}^2$) 
in the denominator in Eq.(34).   For $1+ w_{\rho}n_{\rm s}>0$, 
the   screening length is given by 
\be 
\xi_\rho = \kappa^{-1}\sqrt{1+ w_{\rho}n_{\rm s}}, 
\en 
which  is valid  for $\kappa d_1 \ll  1$ 
or for   $n_{\rm s}\ll 0.02n_{\rm w}\sim 1$ mol$/$L with 
$d_1=3{\rm \AA}$ in water.  In   Sec.IV, 
we shall see that  $w_\rho$ is negative   
and $|w_\rho|$ increases with increasing the cation-anion 
asymmetry (see  Fig.9(d) and Eq.(100)). 
Thus, $\xi_\rho \kappa$ decreases 
with increasing $n_{\rm s}$ for small $n_{\rm s}$ 
as    observed\cite{Perkin}. A similar  decrease was  derived 
in the MSA scheme\cite{Blum1,Blum2} and 
in  phenomenoological theories\cite{Attard,Adar} 
However,  $\xi_\rho$   increases      
with increasing $n_{\rm s}$ above 1 mol$/$L\cite{Perkin,Adar,Coles},  
as a remarkable effect  beyond the scope of  this paper.

In    Eq.(28) 
the cations and the anions are { indistinguishable}. 
Thus,   we   define the Kirkwood-Buff  
integrals (KBIs)\cite{Kirkwood}    
 for  the water  density $n_1$ and the ion density 
$n_{\rm e}= n_2+ n_3$\cite{Patey,Newman,Smith1,Knecht,Netz1,Naleem-Smith}.
Then,  Eqs.(28)-(30)  give  the  ion-ion  and ion-solvent  KBIs:  
\bea
&&\hspace{-2mm}
 G_{\rm ss}=\av{n_{\rm e}:n_{\rm e}}/4n_{\rm s}^2-1/2n_{\rm s} 
=  (2\chi-1)/2n_{\rm s}, \\
&&\hspace{-2mm}
G_{{\rm ws}}=\av{n_1:n_{\rm e}}/2n_{\rm w}n_{\rm s}= 
-\chi(f_{12}+ f_{13})/n_{\rm w}f_{11}. 
\ena  
Thus, as  $n_{\rm s}\to 0$,  we have 
 $ G_{\rm ss}\propto n_{\rm s}^{-1/2}$  and  
  $G_{{\rm ws}}\cong - {v_{\rm s}^*}/{2}$. 
Note that   $G_{{\rm ws}}$ represents exclusion (adsorption) 
 of water molecules  
  around an ion pair for positive (negative) $v_{\rm s}^*$.  

In their simulation, Naleem {\it et al.}\cite{Naleem-Smith}   
found growth of  $G_{\rm ss}$  at low densities  of  CaCl$_2$. 
Here,  we readily derive  
$G_{\rm ss}=  Z_+ Z_- \ell_B \kappa/4\av{n_{\rm e}}+\cdots$,  
 where  the cations and anions 
have changes $Z_+e$ and $-Z_- e$, respectively,

\section{Thermodynamics  of electrolytes}

In this section, we study the electrolyte thermodynamics  
using  the   Helmholtz free energy      
$F= \lim_{\Lambda\to 0}{\cal F}$. 
We give remarks on previous research. 
(i) Pitzer\cite{Pitzer} used   the Gibbs free energy 
$G$. In Appendix B, a scheme of $G$  will be given. 
(ii) In many 
papers\cite{RBook,Bj,DS,Ebeling,Fisher,De,Has,A-B,Roij,MaPair,Vegt3,Fennell},    {\it associated   ion pairs} are  
treated  as dipoles coexisting with unbound ions. 
However,  they appear as  ion clusters  with  finite lifetimes 
in water. In Appendix A,  we  will show how  
our theory is modified by  
 such dipoles at small $n_{\rm s}$. 
(iii)  Since  McQuarrie's paper on fused salts\cite{McQ2}, many 
 authors\cite{Stell,Tomo,Ebeling,Fisher,Card,Pana} discussed  
  a gas-liquid  phase transition of the ions due to  
${f}_{\rm DH}$ in Eq.(5) without solvent-ion interactions, where  
$a_2$ and $a_3$ are  the ion 
 hardsphere diameters.

\subsection{Free energy, chemical potentials, pressure, 
 and thermodynamic derivatives}

From Eqs.(3) and (7)-(9).  $F$ is expressed as 
\bea
 F/ V &&= f_{\rm w}(n_{\rm w})  +  
2 k_B T  [\ln ( n_{\rm s}{\lambda}^3) -1 +\nu(n_{\rm w})] n_{\rm s} 
\nonumber\\
&&  - k_BT   \kappa^{3}/12\pi+  U n_{\rm s}^2 .
\ena 
where  $\lambda=\sqrt{ \lambda_2\lambda_3}$.  
This is the expression 
 up to order $n_{\rm s}^2$.   
 We  introduce  the solvent 
chemical potential $ \mu_{\rm w} $   and  the salt one 
 $\mu_{\rm s}$ (per cation-anion pair) 
from $d(F/V)= \mu_{\rm w} dn_{\rm w}+\mu_{\rm s} dn_{\rm s} $ at fixed $T$. 
The pressure $p = n_{\rm w}\mu_{\rm w} +  n_{\rm s}\mu_{\rm s}-F/V$ 
 satisfies the Gibbs-Duhem relation, 
\be 
dp= n_{\rm w} d\mu_{\rm w}+ n_{\rm s} d\mu_{\rm s}.
\en 
 From  Eq.(38), $\mu_{\rm w} $, $\mu_{\rm s} $, and $p$ are expanded as 
\bea 
&&\hspace{-8mm}
\mu_{\rm w} = \mu^0_{\rm w}(n_{\rm w})   
+ k_BT[2\nu' +\ell_B \kappa\epsilon' / \epsilon]n_{\rm s}
+ U' n_{\rm s}^2 ,\\
&&\hspace{-8mm}
 {\mu}_{\rm s}=  k_BT[2\ln (n_{\rm s}\lambda^3) +2\nu- \ell_B
\kappa]+ 2 U n_{\rm s} , \\
&&\hspace{-8mm} p= p^0_{\rm w}(n_{\rm w})   
+ 2k_BT(1+ n_{\rm w}\nu')n_{\rm s}+(U+ n_{\rm w} U') n_{\rm s}^2 \nonumber\\
&&\hspace{-5mm}  -k_BT \kappa^3
 (1-3 n_{\rm w}\epsilon'/\epsilon)/24\pi  ,
\ena 
where   $U'= \p U(n_{\rm w})/\p n_{\rm w}$. We define 
the chemical potential $\mu^0_{\rm w}= 
\p f_{\rm w}(n_{\rm w})/\p n_{\rm w}$ and the pressure    
$ p^0_{\rm w}= n_{\rm w}\mu^0_{\rm w}- f_{\rm w}$ 
   for  pure solvent  at density $n_{\rm w}$. 
They vary significantly even for a 
small change  of $n_{\rm w}$ 
from   
$n_{\rm w}\p \mu^0_{\rm w}/\p n_{\rm w}= 
 \p p^0_{\rm w}/\p n_{\rm w}= 
1/n_{\rm w} \kappa_{\rm w}=k_BT/\epsilon_{\rm in}$.

Next, the second derivatives of $ F/V$ are written as    
\be 
f_{KM}=\frac{\p^2(F/V) }{\p n_M\p n_K} =\frac{\p\mu_K}{\p n_M}
\quad ( K,M={\rm w}, {\rm s}).
\en 
Here,  $f_{\rm ww}= f_{11}$,   $f_{\rm ws}=\sum_{i=2,3} f_{1i}$, and 
 $f_{\rm ss}= \sum_{i,j=2,3}f_{ij}$ in terms of $f_{ij}$  in Eq.(13).  
Note that  the inverse   
 matrix of $\{f_{KM}\}$ 
is given by $\{  \p n_K/\p \mu_M \}$, where    $n_{\rm w}$ and $n_{\rm s}$ 
 are  functions of $\mu_{\rm w}$ and $\mu_{\rm s}$. 
 The elements of this inverse matrix  are the fluctuation variances among 
$\delta n_1$ and $ (\delta n_2+ \delta n_3)/2$ divided  by $k_BT$. Thus, 
 Eq.(28) gives   
\be 
n_{\rm s} \chi=k_BT(\p n_{\rm s}/\p \mu_{\rm s})_{\mu_{\rm w},T}
=  k_BT/[   f_{\rm ss}-f_{\rm ws }^2/f_{\rm ww}]. 
\en

Let us examine  the isothermal compressibility 
 $\kappa_{T}= -V^{-1} \p V/\p p $, 
where $N_{\rm w}= Vn_{\rm w}$, 
$N_{\rm s}= Vn_{\rm s}$,  and $T$ are fixed in the 
pressure derivative. In terms of $f_{KM}$ in  Eq.(43), 
its inverse is   expressed as    
\bea 
&&\hspace{-1cm} 
 \kappa_T^{-1} =\sum_{K={\rm w}, {\rm s}} n_K \frac{\p p}{\p n_K}= 
 \sum_{K,M={\rm w}, {\rm s}}  n_K n_M f_{KM} \nonumber\\
&&\hspace{-2mm} 
=\kappa_{\rm w}^{-1} + 2k_BT( 1+ 2n_{\rm w}\nu'+ n_{\rm w}^2 
 \nu''  ) n_{\rm s}, 
\ena 
where  the DH term  is of order 
$ n_{\rm s}^{3/2}$ (not written here).  
 For   NaCl in  water,   Millero {\it et al.}\cite{Millero-c} 
found  that 
$({{\kappa_{\rm w}}/{\kappa_T}}-1)/n_{\rm s}$ 
tends to a constant as $n_{\rm s}\to 0$, 
which was $7n_{\rm w}^{-1} $  at $T=303$ K. 
Thus,    $n_{\rm w} \nu''/\nu'  \sim 5$ 
from Eq.(45).

The  thermodynamic   partial volumes are   defined by  
${\bar v}_{K}= (\p V/\p N_{K})_{T,p, N_M}$ 
 ($M\neq K)$. 
Since $V$ and $N_K=Vn_K$ are extensive, 
they satisfy the sum rule   
$ {\bar v}_{\rm w}{n}_{\rm w} +  {\bar v}_{\rm s}{n}_{\rm s}=1$.   
 At fixed $T$, 
the relation   $dV= \sum_{K} {\bar v}_K d N_K - V\kappa_T dp$ then holds 
 yielding  $\kappa_T dp= \sum_{K} {\bar v}_K d n_K $ and  
\be
{\bar v}_{\rm s}=  \kappa_T (\p p/\p n_{\rm s})_{T, n_{\rm w}},  
\quad {\bar v}_{\rm w}=  \kappa_T (\p p/\p n_{\rm w})_{T, n_{\rm s}}.   
\en 
Here,  ${\bar v}_{\rm s}$ is defined  for a cation-anion  pair. 
As $n_{\rm s}\to 0$, Eqs.(42) and (46) give  the infinite-dilution limit, 
\be 
{\bar v}_{{\rm s}}^0=\lim_{n_{\rm s} \to 0}{\bar v}_{\rm s} 
=2\epsilon_{\rm in} ( \nu'+n_{\rm w}^{-1})
=   v_{\rm s}^{*}  + 2k_BT \kappa_{\rm w},  
\en 
where  $ v_{\rm s}^{*} $ appears  in Eq.(21). 
The  difference   ${\bar v}_{{\rm s}}^0 -{v}^*_{{\rm s}}=
  2k_BT \kappa_{\rm w}$  stems  from the ionic  partial  pressure  
$2k_BT n_{\rm s}$ and is 
  $0.12n_{\rm w}^{-1}$ in    ambient  water. 
It is relevant   for small-small ion pairs; for example, 
 $n_{\rm w}{\bar v}_{{\rm s}}^0= -0.09$ 
and $n_{\rm w} v_{\rm s}^{*} = -0.21$  for   NaF. 
The  values of ${\bar v}_{{\rm s}}^0$ 
are  listed for various salts  in the experimental 
reports\cite{Millero,Craig,MilleroTetra}. 
Many authors\cite{Millero,Marcus2011,Hepler,Mukerjee,Padova}
introduced  single-ion   volumes,  which are  
$v_i^*+k_BT \kappa_{\rm w}$ in our notation.

\subsection{Salt-doping and apparent partial volumes  } 

 In  experiments of salt-doping, it follows  an  {\it apparent} partial  volume $v_{\rm s}^{\rm ap}$   from  the  space-filling 
relation\cite{RedlichJPC,RedlichReview,Millero,RBook}:  
\be 
n_{\rm w}/n_{\rm w}^0 + v_{\rm s}^{\rm ap}n_{\rm s}=1,
\en
where   $n_{\rm w}^0$ is the  initial solvent density. 
The salt density is increased from 0 to $n_{\rm s}$. 
The simplest  example is to fix the volume   $V$, 
 where $n_{\rm w}= n_{\rm w}^0$, 
$v_{\rm s}^{\rm ap}=0$, and 
$\ln \gamma_{\pm}= -  \ell_B\kappa/2 + 
 Un_{\rm s}/k_BT$  
from Eq.(41) (see Eq.(54) for the definition of $\gamma_\pm $).

 As a well-known  doping method, let 
  a 1:1 electrolyte region be   in osmotic equiibrium  
with a pure solvent  region, which are separated by 
 a semipermeable membrane\cite{McMillan,Luo,OkamotoOnukiTernary}. 
The    solvent  chemical potential $\mu_{\rm w}$ 
 is commonly given by 
$\mu_{\rm w}(n_{\rm w},n_{\rm s})=\mu_{\rm w}^0(n_{\rm w}^0)$.    
From    $d\mu_{\rm w}=\sum_K f_{\rm wK}dn_{\rm K}=0$, 
we set up  the equation, 
\be 
{dn_{\rm w}}/{dn_{\rm s}} =({\p  n_{\rm w}}/{\p  n_{\rm s}})_{\mu_1,T} 
= -  {  f_{{\rm ws}}}/{ f_{{\rm ww}}}.
\en 
From Appendix C, we find   the apparent 
partial volume,
\be 
v_{\rm s}^{\rm ap}= v_{\rm s}^{*}(n_{\rm w}^0)   
+(\epsilon'/\epsilon)\epsilon_{\rm in}
 \ell_B\kappa+  n_{\rm s} \p  U_{\rm eff}/\p p .
\en   
We also calculate the  osmotic pressure   
 $\Pi= p(n_{\rm w}, n_{\rm s})-p_{\rm w}^0(n_{\rm w}^0)$. 
From  $d\Pi=n_{\rm s} d \mu_{\rm s}$  and Eq.(44), we   find   
\be
 d\Pi/dn_{\rm s}=n_{\rm s}
\sum_K f_{\rm sK}(dn_{\rm K}/dn_{\rm s}) 
= k_BT /\chi, 
\en 
which holds for general $n_{\rm s}$. We integrate Eq.(51) using  Eq.(32) 
to obtain    
\be 
\Pi/2k_BT n_{\rm s}= 1 -\ell_B\kappa/6 
  + U_{\rm eff} n_{\rm s}/2k_BT.
\en

\subsection{Isobaric equilibrium   at fixed $p$}

Most  salt-doping experiments have been performed 
at a constant pressure $p$\cite{Lewis,Bronsted,Ninham1,RBook}. 
In this case,     the  salt  number  is   
 increased  from $0$  to  $N_{\rm s}=Vn_{\rm s}$ with 
\be 
 p(n_{\rm w}, n_{\rm s})= p_{\rm w}^0 (n_{\rm w}^0) ,  
\en 
where  $n_{\rm w}^0 $ is the initial solvent density. 
We can also  fix  the total solvent number 
$N_{\rm w}=V n_{\rm w}=V_0 n_{\rm w}^0$, where 
  $V_0$ is  the initial volume. Then,  
Eq.(48) becomes  $V= V_0+ v_{\rm s}^{\rm ap} N_{\rm s}$. 
This isobaric  $v_{\rm s}^{\rm ap}$ 
has been measured, where 
the product $ \phi_{\rm v}= v_{\rm s}^{\rm ap}N_{\rm A}$ 
   is called    the  apparent  molal volume  with  
  $N_{\rm A}$ being   the Avogadro number. 
In Eq.(B3) in Appendix B, ${\bar v}_{\rm s}$ 
will be expressed in terms of this  $v_{\rm s}^{\rm ap}$.

We define the   {\it molal}  
 mean  activity coefficient $\gamma_{\pm} $ 
by expressing  the salt chemical potential  $\mu_{\rm s}$ in Eq.(41)  as 
\be 
\mu_{\rm s}/2k_BT =  \nu (n_{\rm w}^0) 
+  \ln (\lambda^3\gamma_{\pm} N_{\rm s}/V_0).  
\en  
We  also introduce  the   {\it molar}  
 mean  activity coefficient\cite{RBook}, 
\be 
y_\pm= (V/V_0) \gamma_{\pm}= (1+ v_{\rm s}^{\rm ap}N_s/V_0) \gamma_{\pm}. 
\en   
Then,  $\gamma_{\pm}N_{\rm s}/V_0= n_{\rm s}y_\pm$ in Eq.(54).  
 Setting  $\nu(n_{\rm w}) \cong \nu(n_{\rm w}^0) 
-\nu'n_{\rm w}^0 {\bar v}_{{\rm s}}^0  n_s$ in Eq.(41), we  obtain  
\be 
\ln \gamma_{\pm}= -  \ell_B\kappa/2 + 
 {\tilde U}_{\rm eff} n_{\rm s}/k_BT.
\en 
At fixed $p$ we use the coefficient  $ {\tilde U}_{\rm eff}$  defined by  
\be
{\tilde U}_{\rm eff}= U-({\bar v}_{{\rm s}}^0)^2/2\kappa_{\rm w} =
U_{\rm eff}- k_BT (v_{\rm s}^{*} + {\bar v}_{{\rm s}}^0), 
\en 
where $v_{\rm s}^*$ in Eq.(25) is replaced  by $ {\bar v}_{{\rm s}}^0$ 
in Eq.(47). It 
 will also appear in the Gibbs free energy in Appendix B. 
Note that  ${\tilde U}_{\rm eff}/k_BT$ can be  known from the data of 
$(\ln\gamma_\pm+  \ell_B\kappa/2)/n_{\rm s}$, 
which is slightly negative for LiF 
($\sim -2/n_{\rm w}$)\cite{Hamann}, 
positive for the  the other alkali 
halide  salts, and is largely negative 
 for NaPBh$_4$ ($\sim -60/n_{\rm w}$)\cite{Taylor}. 

We also have   $dp=\sum_{K,M}  (n_K  f_{K M} )  d n_{M}=0$ 
 from Eq.(39).  Using Eq.(43),  we can  set up the equation,      
\be 
\frac{dn_{\rm w}}{dn_{\rm s}}=\ppp{n_{\rm w}}{n_{\rm s}}{p,T} 
 = -  \frac{f_{\rm ws}}{f_{\rm ww} }- 
\frac{ k_BT/\chi}{n_{\rm w}f_{\rm ww}+n_{\rm s}f_{\rm ws}}.  
\en  
Here,  the second term is $(\p n_{\rm s}/\p \mu_{\rm w})_{n_{\rm s},T}
(\p \mu_{\rm w}/\p n_{\rm s})_{p,T}$. 
From Appendix C, we find the aparent partial volume,  
\be 
{v}_{\rm s}^{\rm ap} = {\bar v}_{{\rm s}}^0(n_{\rm w}^0) +
 (\epsilon'/\epsilon-1/3n_{\rm w})\epsilon_{\rm in}
 {\ell}_B  \kappa + h n_{\rm s}.
\en  
The  second  term is the DH part derived by 
Redlich\cite{RedlichJPC,RedlichReview}. 
The   $h$ is  called the deviation 
constant and has been measured 
(see  Table IV in Sec.IV)\cite{Millero,MilleroTetra,Desnoyers}.    
It is expressed as 
\be 
h=  \kappa_{\rm w}{\tilde U}_{\rm eff}+ \p{\tilde U}_{\rm eff}/\p p .  
\en 
We can   also  derive Eqs.(59) and (60)     by 
 expanding $p(n_{\rm w}, n_{\rm s})$ in Eq.(42) 
with respect to $\delta n_{\rm w}= n_{\rm w}- n_{\rm w}^0$.  

The derivative  $d\mu_{\rm w}/dn_{\rm s}=(\p \mu_{\rm w}/\p n_{\rm s})_{p,T}$ 
is given by the second term in Eq.(58) multiplied by $f_{\rm ww}$.   
Its integration gives $\mu_{\rm w}$, leading to 
 $\mu_{\rm w}=  \mu_{\rm w}^0  
 - 2k_BT n_{\rm s}/n_{\rm w}^0+\cdots$ for small $n_{\rm s}$,   
where $\mu_{\rm w}^0=f_{\rm w}'(n_{\rm w}^0)$ 
is the initial  chemical potential of pure solvent. Thus, 
we  define 
\be
 \varphi= [\mu_{\rm w}^0(n_{\rm w}^0) 
 - \mu_{\rm w}(n_{\rm w}, n_{\rm s})]/(2k_BT  n_{\rm s}/n_{\rm w}).
\en 
After some calculations we obtain   the expansion,  
\be
\varphi=1 - \ell_B\kappa/6+ {\tilde U}_{\rm eff}n_{\rm s}/2k_BT .   
\en 
This $\varphi$  is called the osmotic 
coefficient  as well as $\Pi/2k_BTn_{\rm s}$ 
in Eq.(52)\cite{Ninham1,RBook}, but  
  the linear term($\propto n_{\rm s}$)  in Eq.(52) 
is larger  than that in Eq.(62) 
by $ ({\bar v}_{{\rm s}}^0-k_BT\kappa_{\rm w})n_{\rm s}$.   

From ${n_{\rm s}}  {d\mu_{\rm s}}/dn_{\rm s}= 
-{n_{\rm w}} d\mu_{\rm w}/dn_{\rm s}$, 
we also find\cite{Patey}   
\be
1+ n_{\rm s}(\p \ln y_\pm/\p n_{\rm s})_{p,T} 
=1/ [{1+ 2n_{\rm s}( G_{\rm ss}- G_{{\rm ws}})}], 
\en
with the aid of  Eqs.(54) and (55).
 Here,   $G_{\rm ss}$ and $ G_{{\rm ws}}$ are 
 the KBIs in  Eqs.(36) and (37), which  
satisfy  $1+ 2n_{\rm s}(G_{\rm ss}- G_{\rm ws})= 
2\chi(1+ n_{\rm s}f_{\rm ws}/n_{\rm w}f_{\rm ww})$ 
from  Eq.(44).  This relation 
 has been used in simulations
to  calculate   $y_\pm$\cite{Smith1,Knecht,Netz1,Naleem-Smith}.

We make some comments. 
(i) In Appendix B, we will derive  Eqs.(59), (60), and (62)  from 
the Gibbs free energy. 
 (ii)  Bernard {\it et al.}\cite{Ninham1} 
 related  $\varphi$ and $\Pi$  by 
 $\varphi= (1-{\bar v}_{{\rm s}}^0 n_{\rm s})\Pi/2k_BTn_{\rm s}$,  
where   ${\bar v}_{{\rm s}}^0$  should be replaced by 
 ${\bar v}_{{\rm s}}^0-k_BT \kappa_{\rm w}$ in our theory. 
(iii) The behavior $\propto \sqrt{n_{\rm s}}$ of the first  corrections 
in  $\ln\gamma_{\pm}$ and $\varphi$ 
is  the DH limiting law, 
which  was  known empirically   before the DH  theory\cite{Lewis,Bronsted}.

\subsection{Expressions  in  extended Debye-H\"uckel theory}

With increasing $n_{\rm s}$, 
the lowest DH terms    in Eqs.(56), (59), and (62) 
increase    as $\sqrt{n_{\rm s}}$, while  
the Debye length   $\kappa^{-1}$ 
decreases toward  the minimum length  ($a_2$ or $a_3$). 
However, ${f}_{\rm DH}$ in Eq.(5)    is suppressed  with increasing  
$a_i \kappa$. Due to this  reason, 
many authors  used {\it extended} DH expressions 
to explain   experimental  data\cite{RBook,Hamann,Gu1,Gu2,Pitzer}. 

We thus  rewrite  $\gamma_{\pm}$ in Eq.(56)   and 
 $\varphi$ in Eq.(62) as\cite{Gu1,Gu2}     
\bea 
&& \hspace{-10mm}
\ln \gamma_{\pm}= - \frac{1}{4}  \ell_B
\kappa\sum_{i=2,3}  \frac{1}{1+ a_i \kappa}  +b n_{\rm s},\\ 
&&\hspace{-1cm}  \varphi = 
 1- \frac{1}{12} 
\ell_B\kappa\sum_{i=2,3}  \sigma(a_i \kappa)  
 + \frac{1}{2} b'  n_{\rm s},
\ena 
where   $\sigma(x)= 3[x+ x/(1+x)-2\ln (1+x)]/x^3$ and 
  $\sigma(x)=  1-3x/2+\cdots$ for $x\ll 1$. 
For small $a_i\kappa$ we compare Eqs.(64) and (65) and  
  Eqs.(56)  and (62)  to find    
\be
    b= b' \cong {\tilde V}_{\rm eff}/k_BT .
\en 
Here, using ${\tilde U}_{\rm eff}$ in Eq.(57) and 
$u_{ij}^{\rm ex}$ in Eq.(6),  we define  
\be
 {\tilde V}_{\rm eff}={\tilde U}_{\rm eff} -\frac{1}{2}
\sum_{i,j}u_{ij}^{\rm ex}=
{\tilde U}_{\rm eff} -  2\pi k_BT\ell_B^2 (a_2+a_3) ,
\en 
In Appendix D, we will present  extended  DH forms for 
$\chi^{-1}$ in Eq.(32) and $v_{\rm s}^{\rm ap}$ in Eq.(59).

Guggenheim and Turgeon\cite{Gu1,Gu2} nicely 
fitted Eqs.(64) and (65) to  1:1 electrolyte data  
with  $b=b'$ and    $a_2= a_3= 3{\rm \AA}$. 
They used  many data points for each salt. 
For their choice of $a_i$, the relation     
 $a_i\kappa\cong \sqrt{m}$ holds, where    $m$ 
is    the molality.    Many authors\cite{RBook,Hamann,Pitzer} 
 took   this {\it practical} approach  with empirical $b=b'$.

\begin{table}[t]
\caption{  Data of mean activity coefficient $\gamma_{\pm}$ 
and osmotic coefficient $\varphi$ 
for alkali  halide salts at molality $0.5$ in ambient  
water\cite{Hamer}. The latter are in ().   LiF is insoluble at this density.
  } 
\begin{ruledtabular}
\begin{tabular}[t]{c | c c c c}
& $\mathrm{F}^-$   & $\mathrm{Cl}^-$ & $\mathrm{Br}^-$ & $\mathrm{I}^-$\\
\hline
$\mathrm{Li}^+$ &  & 0.739 (0.964) & 0.754 (0.970) & 0.824 (1.008)\\
$\mathrm{Na}^+$ & 0.633 (0.887)& 0.681 (0.921)& 0.697 (0.932)& 0.722 (0.950)\\
$\mathrm{K}^+$ & 0.670 (0.916)& 0.649 (0.900)& 0.658 (0.906)& 0.676 (0.918)\\
$\mathrm{Rb}^+$ & 0.701 (0.939)& 0.633 (0.891)& 0.630 (0.889)& 0.627 (0.887)\\
$\mathrm{Cs}^+$ & 0.721 (0.946)& 0.607 (0.873)& 0.605 (0.870)& 0.601 (0.868)\\
\end{tabular}
\end{ruledtabular}
\end{table}

\subsection{Experimental trends   and Collins' rule}

Table I gives $\gamma_{\pm}$ and  $\varphi$  for alkali  halide salts 
at molality $0.5$ in ambient water\cite{Hamer}. 
 We notice the following.
 (i) For F$^-$, $\gamma_{\pm}$ and $\varphi$ 
 increase  with increasing the  cation size. 
For the other  anions,  they are  smaller for 
larger  cations. (ii) For small cations Li$^+$ and Na$^+$,  
$\gamma_{\pm}$ and $\varphi$    increase 
as the anion size increases. For large cations  Rb$^+$ and Cs$^+$, 
 the tendency is reversed.  
(iii) For K$^+$, they are  close   for all the anions. 
Thus, K$^+$ ions have a marginal  size.

\begin{table}[t]
\caption{ Coefficient  $ b$ in Eq.(64) 
 and ${\tilde V}_{\rm eff}/k_BT $  
from  Eqs.(56) and (67) in units of $d_1^3$. 
The latter are  in ().   
  Use is made of data on   the mean activity coefficient  
for alkali halide salts  at molality 
0.02 for LiF\cite{Hamann} and 0.1 for 
the others\cite{Hamer}.  }
\begin{ruledtabular}
\begin{tabular}[t]{c | c c c c}
& $\mathrm{F}^-$   & $\mathrm{Cl}^-$ & $\mathrm{Br}^-$ & $\mathrm{I}^-$\\
\hline
$\mathrm{Li}^+$ &  -68.0 (-70.4)& 22.1 (11.5) & 28.3 (17.8) & 43.7 (33.2)\\
$\mathrm{Na}^+$ & 2.1 (-8.4) & 14.2 (3.6) & 17.4 (6.79) & 22.1 (11.5) \\
$\mathrm{K}^+$ & 9.4 (-1.2) & 5.4 (-5.2) & 7.8 (-2.8) & 11.8 (1.2) \\
$\mathrm{Rb}^+$ & 15.0 (4.4) & -0.3 (-10.9) & -1.1 (-11.7) & -2.0  (-12.5) \\
$\mathrm{Cs}^+$ & 24.4 (13.9) & -8.5 (-19.1) & -7.7 (-18.3) & -10.2 (-20.7) \\
\end{tabular}
\end{ruledtabular}
\end{table}
 
 Table II gives   $b$ in Eq.(64) 
and    $ {\tilde V}_{\rm eff}/k_BT$ 
  from   Eqs.(56) and (67) in units of $d_1^3=0.9n_{\rm w}^{-1}$, where    $a_2=a_3=3{\rm \AA}$. 
 We  use  data of $\gamma_{\pm}$ 
 at molality 0.02 for LiF\cite{Hamann} and 0.1 for 
the others\cite{Hamer}.  The  molality 0.1 
is not very small with    $a_i \kappa=  0.31$, so  
the numbers of $b$  are larger than those of  $ {\tilde V}_{\rm eff}/k_BT$ 
  by  10. Here, $ n_{\rm w} U_{\rm eff}/k_BT $ is about $ -5$ for LiF 
and is between  $50$ and $110$ for the others. 
These ion-size-dependences are the same  as those  in   Table I.  
In Fig.2, to illustrate this common trend, we plot $b$ in Table II  vs $\alpha_i= 2R_i^{\rm S}/d_1$ 
($i=2$ for cations and $i=3$ for anions) with $d_1= 3~{\rm\AA}$, 
using  the crystal radii $R_i^{\rm S}$ by Shannon\cite{Shannon}. 

Collins\cite{Collins} noticed  the same  pattern in  
the solubility of alkali halide salts in water 
as those in Tables I and II. That is, 
 salts of   large-small pairs are highly soluble, 
whereas salts of  large-large   or
small-small pairs   are much less  soluble.  
In fact,  the solubility   is  
  $0.05$, 1, and 20  mol$/$L for LiF, 
 NaF,  and  LiCl\cite{LiF,Collins}, respectively. 
He argued  that   large-small  pairs 
 remain apart   but cation-anion pairs  with 
 comparable sizes   tend to 
be  closely connected.  Note that 
the salt solubility 
 is correlated with $U_{23}^{\rm eff}$. 

\begin{figure}[tbp]
\begin{center}
\includegraphics[width=220pt]{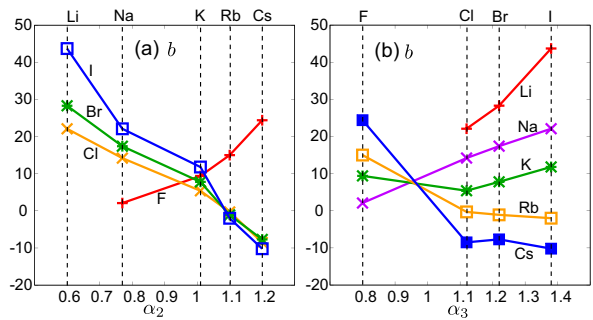}
\caption{(Color online) Coefficient  $ b$ in  
Table II vs radius ratios ($\alpha_2$ for cations in (a)
and  $\alpha_3$ for anions in (b)),  where 
 $b$ appears in the activity coefficient in Eq.(64).   
 Here, Collins' rule holds.  }
\end{center}
\end{figure}

For  NaBPh$_4$\cite{Taylor}, the numbers from the two methods in Table II 
are   $-120.7 ~(-130.5)$ at molality $0.09$,  leading to 
$ U_{\rm eff} \sim {\tilde U}_{\rm eff}   \sim -60k_BT/n_{\rm w} $. 
 For this salt,   the two terms  in Eq.(25) are both about 
$ 1800 k_BT/n_{\rm w}$ from $ v_3^*\cong 15/n_{\rm w}$\cite{MilleroTetra} and 
  their difference  $U_{\rm eff}$ is   much smaller ($\sim 3\%$). 

If the cations and/or the anions are large, 
 $ U_{\rm eff} $ is largely negative from Eq.(25). 
In such cases, a thermodynamic  instability occurs\cite{OkamotoOnukiTernary}   
if $n_{\rm s}$ exceeds a spinodal density 
$n_{\rm s}^{\rm spi}$ determined by    $\chi^{-1}=0$.  For 
$ n_{\rm w}|U_{\rm eff}|/k_BT  \gg  20$ in ambient water, 
the DH term is negligle  in Eq.(32), so      
\be 
 n_{\rm s}^{\rm spi}\sim k_BT/|U_{\rm eff}|. 
\en   
For  NaBPh$_4$, $n_{\rm s}^{\rm spi}$  is    on  
the  order of  its   solubility ($=1.4$ mol$/$L$=0.025 n_{\rm w}$).  
 In  this instability, the ions   aggregate  
as  solvophobic spinodal decomposition\cite{Stell, Onukibook, Glasbrenner}.
  However,   ion association  
can  trigger precipitate  formation 
 in metastable solutions, 
which is the case  for alkali halide  salts in water\cite{aq,Vega,Ya}.   
For LiF,  its solubility ($=0.14$ mol$/$L $=0.002n_{\rm w}$) 
is exceptionally  small ($\ll n_{\rm s}^{\rm spi}$). On the other hand, 
in aqueous mixture solvents,  phase separation can be induced 
even at slight doping of a strongly hydrophilic  
salt\cite{OkamotoOnukiPrec,Onuki1}.

\subsection{Electrostriction from Born theory} 

Let us consider  the  hydration   part of 
the ion chemical potentials due to the ion-dipole interaction, 
written as $k_BT \nu_i^{\rm B}$.
 In the simple   continuum  theory\cite{Born,Millero,Marcus2011}, it is 
the  integral of the electrostatic 
energy density $\epsilon E(r)^2/8\pi$ in the region 
$r>R_i$, where  $E(r)=\pm e/\epsilon r^2$ is the  
electric field  at distance  $r$  
from the ion  and   $R_i$ is called the Born radius. 
Using  the bulk dielectric constant   $\epsilon $, we find 
\be 
k_BT \nu_i^{\rm B}(n_{\rm w}) = (e^2/2 R_i)(1/\epsilon-1)  \quad  (i=2,3),
\en 
where the contribution without polarization is subtracted. 
We assume that $R_i$ is   independent of $n_{\rm w}$, 
while  $\epsilon$ depends on it as in Eq.(17). 
From Eq.(19)    the electrostriction part of  
$v_i^*$  is given by 
\be 
v_i^{\rm B}= \epsilon_{\rm in} d\nu_i^{\rm B}/dn_{\rm w}= 
 - \epsilon_{\rm in }\ell_B  \epsilon'/(2\epsilon R_i), 
\en 
which is rewritten as $(e^2/2R_i)\p\epsilon^{-1}/\p p$, as  
was first derived by Drude and Nernst\cite{Drude}. 
We also assume  homogeneity of   the local  solvent chemical potential 
 around each  ion\cite{Onuki,Landau}.  We then     
  find   the solvent density increase,  
\be 
\delta n_{\rm w} (r) 
= n_{\rm w}^2 \kappa_{\rm w} 
\epsilon' E(r)^2/8\pi= -
n_{\rm w} v_i^{\rm B}R_i /(4\pi r^{4}),    
\en 
whose integration ($r>R_i$) is 
$ -n_{\rm w}v_i^{\rm B}$ as it should be the case. 
In   ambient water, we have $ n_{\rm w} v_i^{\rm B}
\cong -0.24/R_i$ and $\delta n_{\rm w} (r) /n_{\rm w} 
\cong 0.51 /r^4$ with        $R_i$  and 
$r$  in units of $\rm \AA$.
 where $\delta n_{\rm w} (r)$ 
grows   unrealistically   around small ions.

The  Born  expressions are   very approximate.    
In  water,   dielectric saturation occurs 
and  $\epsilon$  nonlinearly  decreases  in  the  immediate  vicinity of 
ions\cite{Padova}$^,$\footnote{The polarization energy of  
a water molecule  around an ion  is 
    $\mu_0 |E(r)|\sim 3k_BT/r^2$ ($r$
in ${\rm \AA}$) outside the hydration shell in ambient water, 
where    $\mu_0=2.3$ D. The polarization saturates 
for $r\ls \sqrt{3}$~${\rm \AA}$.}  
 In fact, Eq.(70) cannot be  well 
fitted to the electrostriction    data\cite{Marcus2011}  
if  $R_i$ is equated with the  radius  calculated 
 from the crystal lattice constants\cite{Shannon}. 
For example,  Mazzini and Craig\cite{Craig} 
 estimated the electrostriction part of 
${\bar v}_{{\rm s}}^0$ in Eq.(47) as $-13.0$ cm$^3/$mol $=-0.72/n_{\rm w}$ 
for  NaCl. This size  is twice as large as 
 that  from  Eq.(70) if we set  $R_2 \sim 1 {\rm \AA}$ for Na$^+$ 
and $R_3 \sim 2 {\rm \AA}$ for Cl$^-$.
Thus, if we use the Born theory with the bulk $\epsilon$ to explain 
the electrostriction data,  
we  should  treat $R_i$  as a short, effective radius 
(see Eq.(86)).

In addition, the static  dielectric 
constant $\epsilon$   depends  on $n_{\rm s}$ 
  as $\epsilon(n_{\rm w}, n_{\rm s}) 
/ \epsilon(n_{\rm w},0)\cong  1- g_1 n_{\rm s}$,  
where  $g_1n_{\rm w} \sim 10$ for alkali 
hallides\cite{VegtPRL,ep-c1,ep-c,Andelman-ep}. 
This indicates that $1/\epsilon$ in Eq.(69) 
should be changed to  $(1+ g_1 n_{\rm s})/\epsilon(n_{\rm w},0)$, 
which yields  an additional positive contribution 
to $U_{\rm eff}$\cite{Boda}. In this paper, we  neglect such an  
indirect repulsive interaction.


\section{Model calculations }

To make numerical analysis, we combine    the MCSL model\cite{MCSL}, 
the attractive part of  the Lennard-Jones (LJ) potentials\cite{Oxtoby}, 
and the Born chemical potentials\cite{Born}. 
Introducing  the hardsphere diameters $d_1$, $d_2$, and $d_3$ 
for the solvent, the cations, and the anions, 
respectively,  we  vary the diameter ratios, 
\be 
\alpha_i= d_i/d_1 .
\en  
The  steric interaction sensitively depends on whether 
$\alpha_2$ and $\alpha_3$ are larger or smaller 
than 1. In the following, large and small 
ions are roughly those with $\alpha_i\gs 1.2$ 
and $\alpha_i\ls 0.8$, respectively. 

\subsection{Local free energy density  $f$}

The   free energy density $ f $ in Eq.(3) is given by    
\be
 f=k_B T\sum_{i=1,2,3}   n_i [\ln( n_i\lambda_i^3)-1]  +f_{\rm DH}  +  f_{\rm h} + f_{\rm a}  +f_{\rm B },  
\en  
where the first term is the ideal-gas part and 
 $f_{\rm DH}$ is the DH free energy density in Eq.(5).
The third  term $f_{\rm h}$ is the MCSL  steric part 
  written up to second order in $n_2$ and $n_3$ as    
\be 
f_{\rm h}= 
 f_{\rm h}^0(n_1) + k_BT \sum_{i=2,3}\nu_i^h n_i +\frac{1}{2} 
\sum_{i,j=2,3}  U_{ij}^hn_i n_j, 
\en 
where $f_h^0$ is given by  the Carnahan-Starling form\cite{Carnahan},  
\be 
f_{h}^0   = k_BTn_1 (4-3\eta_1) \eta_1/(1-\eta_1)^2 ,  
\en 
with  $\eta_1= v_1 n_1$ with $v_1=\pi d_1^3/6$ 
being  the  hardcore volume of a solvent particle.  
 See Appendix E for  expressions of   $\nu_i^h$ and $U_{ij}^h$.  
 The fourth term   $f_{\rm a}$ represents the attractive interaction 
 assuming  the van der Waals  form, 
\be 
f_{\rm a}= - \frac{1}{2}\sum_{i,j=1,2,3} w_{ij} n_i n_j.  
\en  
The coefficients   $w_{ij}$  $(i, j=1,2,3)$ are constants given by 
\begin{align}
w_{ij}= 
({4\sqrt{2}\pi}/{9}) \epsilon_{ij}( d_{i}+ d_j)^3,
\end{align}
where $\epsilon_{ij}$ are  interaction  energies 
  in the   LJ  potentials\cite{Oxtoby}. 
From Eq.(69)  the hydration part 
 $f_{\rm B}$ is  written   as  
\be 
 f_{\rm B} = k_BT \sum_{i=2,3} \nu_i^{\rm B} (n_1)  n_i.
\en

The free energy density of pure solvent is given by\cite{CarnahanE} 
\be 
f_{\rm w}(n_1) = k_BT n_1(\ln (n_1\lambda_1^3)-1] + f_{h}^0(n_1)  - 
\frac{1}{2}w_{11} n_1^2.  
\en 
The  incompressibility  parameter   
 $\epsilon_{\rm in}$  in Eq.(10) becomes   
\be
\epsilon_{\rm in} =[1/\epsilon_{\rm in}^h  -n_1 {w_{11}}/{k_BT}]^{-1},
\en 
where $\epsilon_{\rm in}^h $ is the hardcore part. 
Its iverse is written as\cite{Carnahan} 
\be  
1/\epsilon_{\rm in}^h = 1+{2\eta_1(4-\eta_1)}/{(1-\eta_1)^4}, 
\en 
where the second term grows for $\eta_1\gs 0.5$. 
 For water,  the hydrogen bonding  yields 
a  high critical temperature ($647.1$K), so  we need a relatively  
large  $w_{11}$ to make  the phase diagram from    $f_{\rm w}$  mimic that of 
water\cite{OkamotoOnukiTernary}. Thus, we  introduce 
 the attraction  parameter of the solvent,     
\be 
w_{\rm a} = \epsilon_{\rm in}/\epsilon_{\rm in}^h-1
= \epsilon_{\rm in} n_1 w_{11}/k_BT , 
\en 
which is of order 1 for ambient water as its speciality.

We set $d_1$ and $\epsilon_{11}$ in  $f_{\rm w}$ in Eq.(79) equal to    
\begin{align}
&d_1=3\, \mathrm{\AA}, \quad \epsilon_{11}/k_B =412.72\, \mathrm{K}.
\end{align} 
For  ambient water ($T=300\,\mathrm{K}$ and $p=1\,\mathrm{atm}$), 
these   give  the experimental compressibility 
$\kappa_{\rm w}=4.5\times 10^{-4}$ {MPa}$^{-1}$. 
We also  obtain   $n_1 = 0.857/d_1^3= 31.7${nm}$^{-3}$, which  is slightly smaller than the experimental one 
$=33.3$~{nm}$^{-3}$. 
Then, $1/\epsilon_{\rm in}^h=   35.5$ and  $n_1 {w_{11}}/k_BT= 18.6$. Thus,  
\be 
\eta_1=v_1n_1= 
0.448, \quad 
\epsilon_{\rm in}=0.059, \quad   
w_{\rm a} =1.10.
\en  
Previously\cite{OkamotoOnukiDFT,OkamotoOnukiTernary,OkamotoOnukiBubble},  
we assumed  $\epsilon_{11}/k_B= 588.76$ K to obtain   
the  saturated vapor pressure of water ($=0.031$ atm)  at $T=300$ K. 
As regards the dielectric constant, we set  $\epsilon=80$ 
and $n_1\epsilon'/\epsilon=1.1$ in accord with    Eq.(17).

  The other  LJ   energies in Eq.(77) are  given by    
\be 
 \epsilon_{1i}/k_B= 287.3, \mathrm{K},~  
\epsilon_{ij}/k_B=200,\mathrm{K}~~ (i,j=2,3), 
\en
which are smaller than $\epsilon_{11}/k_B$ and satisfy 
 the Lorentz-Berthelot relations\cite{Hansen}  
 $\epsilon_{ij}=\sqrt{\epsilon_{ii}\epsilon_{jj}}$. 
For simplicity, we set $\epsilon_{12}=\epsilon_{13}$ 
 not differentiating  the properties of cations and anions in water, 
so we can   exchange $\alpha_2$ and $\alpha_3$  in our results. 
 In molecular dynamics simulation of aqueous electrolytes\cite{aq,Vega,Ya}, 
the pair potentials  among  ions and water molecules 
depend on the ion species.  

As discussed in Sec.IIIF, to be consistent 
with the electrostriction data, the  Born radii $R_i$  
should be  smaller than  the hardsphere radii  $d_i/2$. 
In this paper, we set   
\begin{align}
R_i=0.2 d_i \quad (i=2,3). 
\end{align} 
Then,  we have $v_i^*<0$ for $\alpha_i<0.72$ (see Fig.3(a)). 
If   $R_i=0.4d_i$, we have  $v_i^*<0$ for $\alpha_i<0.58$.

\subsection{Ion volume and interaction coefficients}

The  solvation coefficient $\nu_i(n_1)$  in Eq.(3) consists of 
three parts as     
$\nu_i(n_1) = \nu_i^h - w_{1i} n_1/k_BT + \nu_i^{\rm B}$.
Then, from Eq.(19),    the ion volume is written as   
\be 
v_i^*=  v_i^h+v_i^{\rm LJ}  + v_i^{\rm B}.  
\en    
The   MCSL part $v_i^h=\epsilon_{\rm in} d\nu_i^h/d n_1$ 
tends to  $v_1 \alpha_i^3$ for large $\alpha_1 $   
(see  Eq.(E4) in Appendix E for its expression). With  Eqs.(83)-(86), 
the LJ  part  $v_i^{\rm LJ} =- {\epsilon_{\rm in} w_{1i}}/{k_BT}$ and the 
Born part $v_i^{\rm B}$  in Eq.(70)   behave as 
\be
 v_i^{\rm LJ}/d_1^3= -0.11 (1+\alpha_i)^3 , \quad 
v_i^{\rm B}/d_1^3= -0.44/\alpha_i.
\en 

\begin{figure}[tbp]
\begin{center}
\includegraphics[width=245pt]{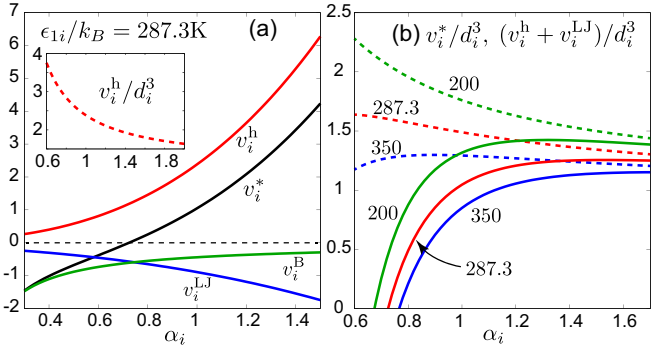}
\caption{(Color online) (a) Infinite-dilution ion volume 
$v_i^*$ (black) composed of  $ v_i^h$ (red), $ v_i^{\rm LJ}$ (blue), 
and $ v_i^{\rm B}$ (green) 
 in units of $d_1^3$ together with 
  $v_i^h/d_i^3$ (inset)   as functions of $\alpha_i= d_i/d_1$. 
(b) Ratios $v_i^*/d_i^3$ (bold line) 
and $(v_i^h+ v_i^{\rm LJ}) /d_i^3$ (broken line) for  
 $\epsilon_{1i}/k_B = 200, 287.5$, and $350$K.  
In the other figures,  $\epsilon_{1i}/k_B =  287.5$K.  
 }
\end{center}
\end{figure}

In Fig.3(a),  we examine the three  ion-volume parts. For  
 $\alpha_i\ls 0.5$, we have $v_i^*\sim  v_i^{\rm B}<0$.
For  $\alpha_i > 1$,  both   $v_i^h $  and $v_i^{\rm LJ}$  
grow as $\alpha_i^3$,  where  $v_i^{\rm B}$ is negligible. 
 In (b),  we plot  the ratios 
$ v_i^*/d_i^3$ and $ (v_i^h+v_i^{\rm LJ})/d_i^3$ for  
  $\epsilon_{1i}/k_B =$200,  287.5, and $350$K, which  
  decrease with increasing $\epsilon_{1i}$.  For $\alpha_i\gs 1.2$, 
we can neglect $v_i^{\rm B}$ and find 
\be 
{v_i^* }\cong {v_1}   (1+ w_{\rm a})\alpha_i^3. 
\en 
 See Eq,(E4) and the sentences below it.

To understand the overall behavior of $v_i^*$,  we give 
 a simple interpolation formula, 
\be 
v_i^*/d_1^3  \cong   D_{\rm L}  \alpha_i^3- D_{\rm B} /\alpha_i ,
\en  
Here,  $D_{\rm L}=\pi(1+w_{\rm a})/6 = 1.1$  from Eq.(89) 
and  $D_{\rm B}=\epsilon_{\rm in}\ell_B 
\epsilon'/(0.4d_1^4\epsilon)=0.44$ from Eqs.(70) and (86). 
If   $v_i^*=0$,  Eq.(90)  yields  $\alpha_i=0.80$, 
while our  full equations give  $\alpha_i=0.72$ in Fig.3(a).  
Previously, some authors\cite{Hepler,Mukerjee,Padova,Millero,Marcus2011} 
  wrote   the ion volume  ($=v_i^* + k_BT \kappa_{\rm w}$)  
 in  the form  $A_{\rm I}(2r)^3- B_{\rm I} /r$, where $r$ is 
a certain  ion radius 
with $A_{\rm I}$ and $ B_{\rm I}$ being constants.
  They set   $A_{\rm I}\cong 1.0$ in agreement with 
our  $D_{\rm L}=1.1$ (if their $r$ is assumed to be 
close to   the crystal radius).


We next show  the salient features of 
the interaction coefficients.   
From Eq.(5)  $U_{ij}$  in Eq.(3) 
and $U_{ij}^{\rm eff}$  in Eq.(24)  include 
 $u_{ij}^{\rm ex}$ in Eq.(6). We calculate  the  excess parts,   
\bea 
&& V_{ij}=U_{ij} - u_{ij}^{\rm ex} ,\quad 
V_{ij}^{\rm eff}=U_{ij}^{\rm eff} - u_{ij}^{\rm ex}, \nonumber\\
&& V_{\rm eff}=  U_{\rm eff}-  \frac{1}{2}\sum_{i,j}
u_{ij}^{\rm ex}.    
\ena 
We have introduced ${\tilde V}_{\rm eff}$ in Eq.(67). 
From Eq.(73)  $V_{ij}$ consist of 
the MCSL and LJ parts as $V_{ij}=U_{ij}^h - w_{ij}$.

\begin{figure}[t]
\begin{center}
\includegraphics[width=225pt]{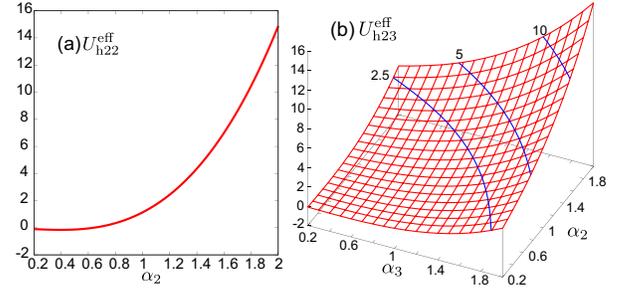}
\caption{(Color online) (a) 
${U_{h 22}^{\rm eff} }$ 
vs $\alpha_2$ and (b) ${U_{h 23}^{\rm eff} }$ 
in the $\alpha_2$-$\alpha_3$ plane in units of ${d_1^3 k_BT}$, 
where  $\eta_1=0.448$ 
 and $\epsilon_{\rm in}^h=0.028$. 
These are the effective interaction coefficients 
in Eq.(92) for purely steric hardsphere mixtures in the MCSL model. 
 }
 \label{UeffHS}
\end{center}
\end{figure}

We consider  the purely steric hardsphere parts of $U_{ij}^{\rm eff}$:
\be 
U_{hij}^{\rm eff}= U_{ij}^h - k_BT n_1\epsilon_{\rm in}^h
(d\nu_i^h/d n_1)(d\nu_j^h/d n_1),
\en  
which will be explicitly calculated  in  Appendix E. 
 In  Fig.~\ref{UeffHS}, 
we display  ${U_{h 22}^{\rm eff} }/d_1^3k_BT$ 
 and  ${U_{h 23}^{\rm eff} }/d_1^3k_BT$. 
The former depends on  $\alpha_2$ only,  
being nearly zero for $\alpha_2<1$ and about 15   
for $\alpha_2\sim 2$. 
The latter is  nearly zero for $\alpha_2<1$ and  $\alpha_3<1$ 
and are about  10 for $\alpha_2\sim \alpha_3\sim 1.8$.  
The two terms in Eq.(92) are both of order $1200d_1^3k_BT$ 
for  $\alpha_2\sim \alpha_3\sim 1.8$, so they 
largely cancel. Thus,  $U_{hij}^{\rm eff}$ 
are smaller than the other contributions with significant 
attractive and hydration interactions. 

Neglecting  $U_{h ij}^{\rm eff}$ in  $V_{ij}^{\rm eff}$, 
we find some simple limiting 
behaviors.  If $\alpha_2$ and $\alpha_3$ are both large, 
we obtain 
 \be 
V_{ij}^{\rm eff}/k_BTv_1 \cong  -\alpha_i^3\alpha_j^3
 (w_{\rm a} +w_{\rm a}^2)/\epsilon_{\rm in}, 
\en 
which are largely negative since $\epsilon_{\rm in}\ll 1$.  
Thus,  salts with large-large ion pairs are 
hardly soluble in water. 
 This is   related to  the hydrophobic assembly 
in water, which has been discussed for uncharged  
large particles\cite{Chandler}. 
Furthermore, if  $\alpha_2$ is small and $\alpha_3$ is large, 
we  obtain 
\be
V_{23}^{\rm eff}/k_BT v_1
\cong  -  v_2^* \alpha_3^3(1+w_{\rm a}) n_1/\epsilon_{\rm in} ,
\en 
which is largely positive for $v_2^* <0$. 
Such asymmetric salts  are considerably  soluble in water\cite{Collins}.

 The cancellation of the two hrdsphere  parts in Eqs.(24) and (92) 
is a general feature. It   is already  indicated by  
the  $\gamma_{\pm}$-data    of  NaBPh$_4$\cite{Taylor} (see  Sec.IIIE).   
 For a neutral solute, Cerdeiri$\rm{\tilde{n}}$a and
  Widom\cite{Widom2}   calculated    the  two terms  in Eq.(1)  with  a smaller difference (see their Fig.3).

\begin{figure}[tbp]
\begin{center}
\includegraphics[width=240pt]{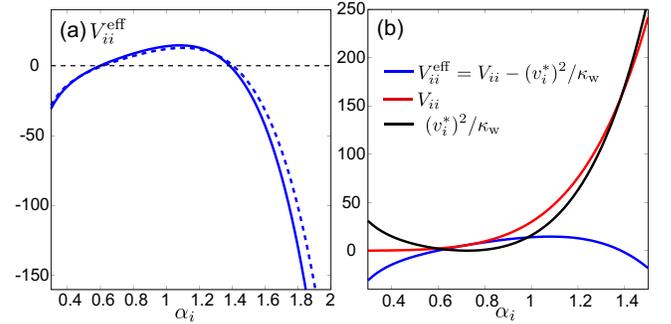}
\caption{(Color online)  
(a) Diagonal (cation-cation or 
anion-anion) component $V_{ii}^{\rm eff}= 
V_{ii}- (v_i^*)^2/\kappa_{\rm w}$  in Eq.(91) 
 vs  $\alpha_i$ in units of $k_BTd_1^3$. 
Its approximation in Eq.(99) is also plotted 
(broken line). 
(b) Comparison of  $V_{ii}^{\rm eff}$ (blue), $V_{ii}$ (red),  and 
$(v_i^*)^2/\kappa_{\rm w}$ (black). For large $\alpha_i$, 
the latter two parts grow but mostly cancel, leading to 
negative $V_{ii}^{\rm eff}$. }
 \end{center}
\end{figure}
\begin{figure}[tbp]
\begin{center}
\includegraphics[width=240pt]{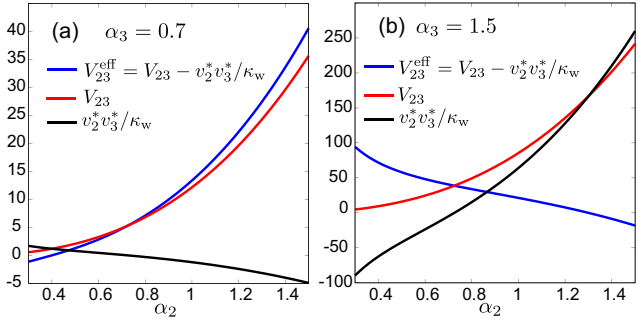}
\caption{(Color online) Off-diagonal (cation-anion) components  
 $V_{23}^{\rm eff}$ (blue), $  V_{23}$ (red), and $  v_2^*v_3^*/\kappa_{\rm w}
$ (black)  in units of $k_BTd_1^3$ as functions of $\alpha_2$, 
where   $\alpha_3$ is (a) $ 0.7$  and (b) $ 1.5$.  
Here,  $  v_2^*v_3^*/\kappa_{\rm w}$ is relatively small in (a), while 
it is largely negative for  $\alpha_2<0.6$ 
and largely positive for $\alpha_2>1.2$ in (b).   
  }
\end{center}
\end{figure}

\subsection{Numerical results of $V_{ij}^{\rm eff}$ }

\begin{figure}[tbp]
\begin{center}
\includegraphics[width=230pt]{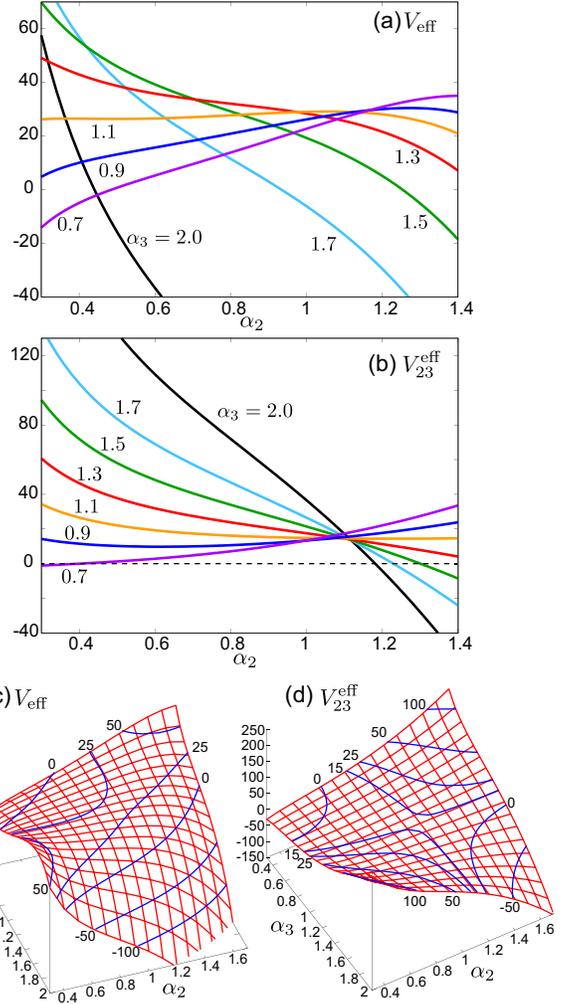}
\caption{(Color online) (a) $V_{\rm eff}$ 
 and (b) $V_{23}^{\rm eff} $  in units of 
$k_BTd_1^3$  vs $\alpha_2$  for $\alpha_3=0.7$, $0.9$, $1.1$, $1.3$,  $1.5$, 
and $2.0$.  Displayed also are bird-eye views 
of  (c)  $V_{\rm eff}$ 
and  (d) $V_{23}^{\rm eff} $ in the region 
$0.35<\alpha_2<1.7$ and $0.35<\alpha_3<2$. They are negative 
for small-small and large-large pairs, but are  
positive  for small-large pairs (see two peaks)\cite{Collins}. 
In (a),  the curves  decrease into negative 
regions rapidly for $\alpha_3>1.5$ 
due to     diagonal  $V_{ii}^{\rm eff}$. The lines of  $\alpha_3=1.1$ 
in (a) and (b) are nearly horizontal in the displayed range, which 
correspond to the contour lines of height 25 in (c) 
and  height 15 in (d). 
 }
\end{center}
\end{figure}

We present some numerical results. In Fig.5(a),  the diagonal component  
$V_{ii}^{\rm eff}$ in Eq.(91) is plotted vs $\alpha_i$, which 
 is independent  of  $\alpha_j$ ($j\neq i$).
It is  positive in the range   $0.60<\alpha_i<1.38$ 
and  is negative outside it   decreasing as  
 $-{\rm const.}\alpha_i^{6}$ for $\alpha_i>1.5$. 
We also plot its approximation to be presented in Eq.(98). 
In (b), we  plot $V_{ii}^{\rm eff}$, 
$V_{ii}$, and $ (v_i)^2/\kappa_{\rm w}$ 
 vs $\alpha_i$. For   $\alpha_i> 1.2$, 
the latter  two  are large and close. 
For  $\alpha_i< 0.5$, we have $V_{ii}^{\rm eff}
\cong - (v_i^{\rm B})^2/\kappa_{\rm w}$.

In Fig.6, we show the off-diagonal components  
$V_{23}^{\rm eff}$,  $V_{23} $, and  
$ v_2^*v_3^*/\kappa_{\rm w}$ vs $\alpha_2$ at fixed  $\alpha_3$.  
Here, $ v_2^*v_3^*/\kappa_{\rm w}$ 
   behaves  very differently for  (a) 
 $\alpha_3=0.7$ and (b) $\alpha_3=1.5$  changing 
 its sign   at  $\alpha_2=0.72$. 
In (a),  $V_{23}^{\rm eff}$ and $V_{23}$ are  close 
and monotonically increase  with increasing  $\alpha_2$, 
where $V_{23}^{\rm eff}=0$ at $\alpha_2=0.40$. 
In (b),  $V_{23}^{\rm eff}$   monotonically 
decreases with increasing  $\alpha_2$ and is negative 
for   $\alpha_2> 1.30$, 
where   $V_{23} $ and $ v_2^*v_3^*/\kappa_{\rm w}$ largely cancel.

In Fig.7,     we display  
   $V_{\rm eff} $  and $V_{23}^{\rm eff} $ 
as functions of  $\alpha_2$    and  $\alpha_3$, 
whose  behaviors change   abruptly  
  as  $\alpha_2$ or $\alpha_3$ changes across  1. 
(i) They   are largely positive for 
 $\alpha_2<1<\alpha_3$ or  $\alpha_3<1<\alpha_2$.   
but  are  negative if both $\alpha_2$ and $\alpha_3$, are large or small. 
 The  $V_{\rm eff} $      is mostly close to 
 ${\tilde V}_{\rm eff} $ in Eq.(67). 
(ii) They increase (decrease)  
with increasing $\alpha_2$  for  small $\alpha_3<1$ 
(large $\alpha_3>1$). See the same tendency 
 in  Table II and Fig.2 for  alkali hallide salts.  
(iii) The lines of $\alpha_3=1.1$ in (a) and (b) 
are  nearly horizontal in the displayed  $\alpha_2$ range. This 
explains  the marginal behavior of K$^+$. 
In  Appendix F, we will explain mathematically why 
  $V_{\rm eff} $  and $V_{23}^{\rm eff} $ 
change their dependence on $\alpha_2$ 
 at $\alpha_3\sim 1$.

\begin{table}[tbh]
\caption{ Example of interaction coefficients  $V_{ij}$, $V_{ij}^{\rm eff}$, 
and $V_{\rm eff}$  in units of  $k_BT d_1^3$ for small-large 
ion pair.
}
\begin{ruledtabular}
\begin{tabular}[t]{c c c  c c c c c c}
$\alpha_2$ & $\alpha_3$ & $V_{22}$ & $V_{22}^{\rm eff}$ 
  & $V_{23}$ & $V_{23}^{\rm eff}$ 
& $V_{33}$ & $V_{33}^{\rm eff}$ & $V_{\rm eff}$\\
\hline 
0.7 & 2 & 
4.86 & 4.76 &  78.2  & 89.6 & 1108  & -291 & -53.8\\
\end{tabular}
\end{ruledtabular}
\end{table}

Table III   gives      $V_{ij}$, $V_{ij}^{\rm eff}$, 
and $V_{\rm eff}$ for $(\alpha_2,  \alpha_3)= (0.7,2)$, 
where $v_2^*/d_1^3=  -0.080 $,  and $ v_3^*/d_1^3=  9.83$. 
In this case,    $V_{33}$  and $(v_3^*)^2/\kappa_{\rm w}$ 
are very  large and close, leading to   
 $V_{23}^{\rm eff}\sim 90$ and $V_{\rm eff}\sim -50$ 
in units of $k_BT d_1^3$. 
For  NaBPh$_4$, we expect similar  behavior 
(see  Sec.IIIE).

\begin{figure}[tbp]
\begin{center}
\includegraphics[width=240pt]{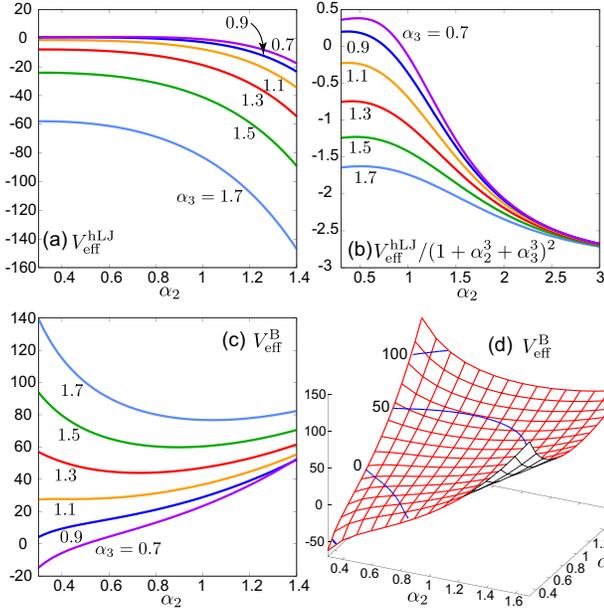}
\caption{(Color online) 
 (a) Non-Born coefficient $V_{\rm eff}^{\rm hLJ}$ in Eq.(96), 
(b) $V_{\rm eff}^{\rm hLJ}/(1+\gamma^2)$ 
with $\gamma=  \alpha_2^3+\alpha_3^3$, and (c) Born coefficient 
$V_{\rm eff}^{\rm B}$ in Eq.(95) as functions of 
$\alpha_2$ for $\alpha_3=0.7,  0.9,1.1, 1.3,1.5$, and $1.7$. 
Shown in (d) is  $V_{\rm eff}^{\rm B}$ in the 
$\alpha_2$-$\alpha_3$ plane. 
These are in units of $k_BTd_1^3$.
   }
\end{center}
\end{figure}

\subsection{Role  of hydration for small-large pairs} 

As in  Eq.(94),  the interplay of the steric and hydration effects 
leads  to the unique behavior of small-large ion pairs. 
  In  $V_{\rm eff}$ in Eq.(91), it give rise to     
\be 
V_{\rm eff}^{\rm B}=|v_{\rm B}|
 (2v_{\rm s}^* + |v_{\rm B}|) /2\kappa_{\rm w}
\en 
where $v_{\rm B}= v_2^{\rm B}+v_3^{\rm B}<0$. 
 We then define the non-Born coefficients without hydration as 
\be 
V_{\rm eff}^{\rm hLJ}=V_{\rm eff}- V^{\rm B}_{\rm eff}.
\en    
In Fig.8, we examine   $V_{\rm eff}^{\rm B}$ and $V_{\rm eff}^{\rm hLJ}$.  
In (a) and (b),   $V_{\rm eff}^{\rm hLJ}$  
is largely negative for $\gamma=   \alpha_2^3+\alpha_3^3>1$ 
and   is small for $\gamma<1$. It is simply   approximated by 
   $V_{\rm eff}^{\rm hLJ} /d_1^3k_BT \cong  
-A\gamma^2/2$ with $A=5.0$. On the other hand, 
in (c) and (d), $V_{\rm eff}^{\rm B}$   is largely positive   for small-large 
pairs and is negative for small-small pairs.

We can  devise a simple approximate expression for 
$V_{\rm eff}$ in terms of $\gamma=   \alpha_2^3+\alpha_3^3$ 
and $\zeta=  1/\alpha_2+1/\alpha_3$ as  
\be 
V_{\rm eff}/d_1^3k_BT \cong   B \gamma \zeta - C\zeta^2/2  -A \gamma^2/2,
\en 
where we use  Eq.(90).  Here, 
 $B=D_{\rm B}D_{\rm L} d_1^3n_1/\epsilon_{\rm in}= 7.0$,   
 $C=BD_{\rm B}/D_{\rm L}= 2.8$, and  $A=5$. 
In the same manner, we express  the cmponents $V_{ii}^{\rm eff}$ and 
 $V_{23}^{\rm eff}$  as 
\bea
&&\hspace{-5mm}   V_{ii}^{\rm eff}/d_1^3k_BT \cong 
2B\alpha_i^2 -C/\alpha_i^2  - A \alpha_i^6 ,\\
&&\hspace{-1cm} 
V_{23}^{\rm eff}/d_1^3k_BT \cong 
[B(\alpha_2^4+ \alpha_3^4) -C]/\alpha_2\alpha_3 
 - A \alpha_2^3\alpha_3^3 .
\ena 
These simple  expressions  can well describe the overall behaviors of 
$V_{ij}^{\rm eff} $ in Figs.5-7.


\subsection{Numerical results of $\chi^{-1}$, $w_\rho$, $G_{\rm ee}$, 
$\ln\gamma_{\pm}$, and  $\varphi$}

\begin{figure}[tbp]
\begin{center}
\includegraphics[width=235pt]{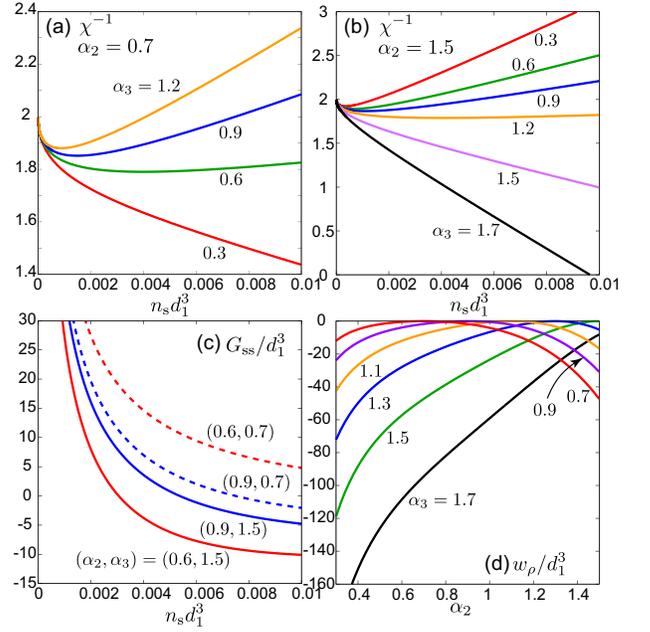}
\caption{(Color online)  $\chi^{-1}$ vs $n_{\rm s}d_1^3$ 
  for $\alpha_3= 0.3, 0.6, 0.9$, and 1.2,  
 where  $\alpha_2$ is (a) $0.7$ and  (b) 1.5 
and use is made of Eq.(D1).
(c) $w_\rho d_1^3$ in Eq.(33) vs $\alpha_2$ for 
 $\alpha_3= 0.7 +0.2m$ ($0\le m\le 5$),
(d) $G_{\rm ss}/d_1^3$ in Eq.(36) vs $n_{\rm s}d_1^3$.  
 for  ($\alpha_2, \alpha_3)=(0.6,0.7), (0.9,0.7),(0.9,1.5)$, 
and ($0.6,1.5)$.  Here, the upper bound of the  salt density 
 $n_{\rm s}$ is $0.01/d_1^3\sim 0.5$ mol$/$L. 
 }
\end{center}
\end{figure}

In Fig.9, setting   (a) $\alpha_2=0.7$  
and (b) 1.5, we plot $\chi^{-1}$   vs $n_{\rm s}d_1^3$($=0.86n_{\rm s}/n_1 $) 
 for various $\alpha_3$,  We use 
  its extended DH form (D1) with $a_2= a_3=3~{\rm \AA}$, 
where $n_{\rm s} \chi$ represents 
the  ionic fluctuation variances    in Eq.(28). 
The coefficient of its linear term     
 $2V_{\rm eff}/k_BT$ 
  is   negative for small-small ion pairs in (a) 
and  large-large  ion pairs in (b). 
For $(\alpha_2, \alpha_3)=(1.7,1.5)$ in (b), 
$\chi^{-1}$ even decreases to 0, resulting in  the instability  
 discussed around   Eq.(68).
 In (c), we also plot the ion-ion KB integral $G_{\rm ss}$ in Eq.(36) 
vs $n_{\rm s}$ for four sets of $(\alpha_2,\alpha_3)$. 
It  grows as $n_{\rm s}^{-1/2}$ as $n_{\rm s}\to 0$.   

In Fig.9(d), we  show  $w_\rho $ in Eq.(33) vs $\alpha_2$ for 
 various $\alpha_3$, which is nonpositive,   
vanishing for $\alpha_2=\alpha_3$. From Eqs.(98) and (99),  
we obtain its approximation, 
\be 
\frac{w_\rho}{d_1^3}= -B(\alpha_2-\alpha_3)^2
\Big[\frac{\alpha_2}{\alpha_3}+\frac{\alpha_3}{\alpha_2}+1 
+\frac{C/2B}{(\alpha_2\alpha_3)^2}+ \frac{A}{2B} \Big].
\en    
Here,  we use  Eq.(85). For general $\epsilon_{ij}$, 
Eq.(33) gives  $w_\rho= (2w_{23}-w_{22}-w_{33})/2k_BT$ 
at  $\alpha_2=\alpha_3$ in   the MCSL model. 
In particular, $w_\rho$ is largely negative 
for large-small ion pairs with $\alpha_2<1<\alpha_3$, 
 for which  ${w_\rho}/ {d_1^3}
\cong  -7\alpha_3^3/\alpha_2$.

In Fig.10, we plot   $\ln\gamma_{\pm}$  
 and   vs   $n_{\rm s}d_1^3$    
for  various $\alpha_2$ and $\alpha_3$. We  use  the extended DH 
expressions (64)-(66)  with $a_2= a_3=3~{\rm \AA}$. 
 These curves are above (below) DH limiting  ones  if  
${\tilde U}_{\rm eff}$ in Eq.(57) is positive (negative) 
 from Eqs.(56) and (62).  
In (a) and (c),   ${\tilde U}_{\rm eff} = -5.0 k_BT d_1^3$  
for $(\alpha_2,\alpha_3)=(0.7,2)$.
Many authors displayed   
 $\ln\gamma_{\pm}$ and  $\varphi$ for   salts 
with  positive linear coefficients\cite{RBook,Hamann,Gu1,Gu2}.

\begin{figure}[tbp]
\begin{center}
\includegraphics[width=245pt]{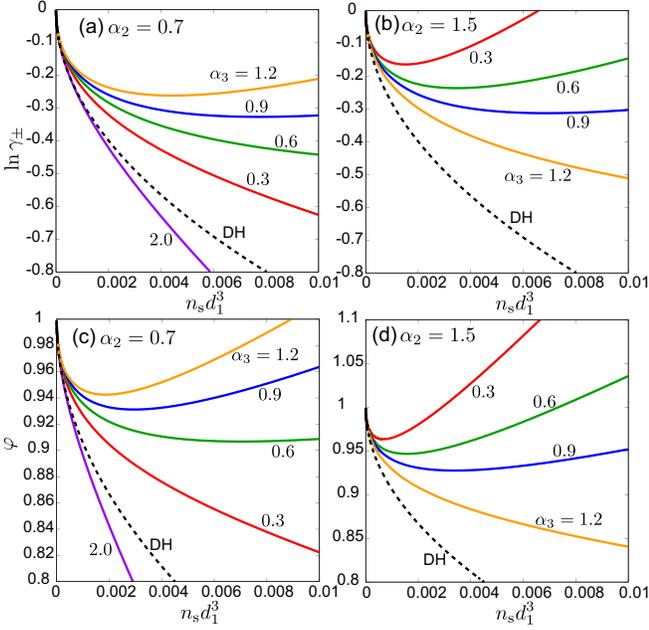}
\caption{(Color online)  
 $\ln\gamma_{\pm}$ and  $\varphi$ in Eqs.(64)-(66) 
 as functions of $n_{\rm s}d_1^3$ 
for  $\alpha_3= 0.3, 0.6, 0.9$, and $1.2$, where  $\alpha_2$ is   $0.7$ in (a) and (c) and is  1.5 in (b) and (d).  The  DH limiting curves 
are also shown (broken lines). 
 }
 \label{fig_activity}
\end{center}
\end{figure}

\subsection{Deviation constant  $h$}

Finally, we examine the deviation constant $h$ 
in   the apparent partial volume $v_{\rm s}^{\rm ap}$ in 
Eq.(59)\cite{Desnoyers,DesnoyersT,Millero,MilleroTetra}. 
Experimentally,  the ion-size-dependence of 
$h$ is opposite to that of   $\ln\gamma_{\pm}$ and $\varphi$, as shown in  
 Table IV.  (i) We  first consider alkali halides\cite{Desnoyers}. 
 For  ${F}^-$, $h$ decreases as the cation size increases. 
For the other anions,  it exhibits the reverse  dependence on the cation size. 
On the other hand, for  cations of not large  size (Li$^+$, Na$^+$, and K$^+$), $h$  decreases as the  anion size increases. 
For  large  {Rb}$^+$ and {Cs}$^+$, $h$ behaves   non-monotonically. 
(ii) Second, for tetraalkylammonium  Et$_4$N$^+$ halides\cite{DesnoyersT}, 
  $h$ is negative and increases  with increasing the anion size.

In our scheme, the unique behavior  of $h$  arises if    
  $\p {\tilde V}_{\rm eff}/\p p$ exceeds 
 $\kappa_{\rm w} {\tilde V}_{\rm eff}$ 
 in Eq.(60).  In particular,  $v_i^{\rm B}$ 
depends on   $n_1$, so  we consider    the ratio 
$A_{\rm B} = n_1 (\p  v_i^{\rm B}/\p n_1) /v_i^{\rm B}$. 
From Eq.(70) it  is expressed as  
\be 
A_{\rm B} = 
(\p^2 \epsilon/\p p^2)/(\epsilon \kappa_{\rm w}^2 a_\epsilon)-2a_\epsilon,
\en   
where    $a_\epsilon= n_1\epsilon'/\epsilon= 1.1$ in  Eq.(17) and $R_i$ 
is assumed to be independent of $n_1$. 
Here, data of $\epsilon$ in ambient water\cite{Archer,Sengers} 
give  $(\p^2 \epsilon/\p p^2)_T \sim   -6\times 
10^{-7}/$MPa$^2$. Thus,  we estimate  $A_{\rm B}\sim -5$.

In Fig.11, we plot  $h$ and  $\p {\tilde U}_{\rm eff}/\p p$  
vs $\alpha_2$ for various   $\alpha_3$ setting 
  $A_{\rm B}=-7.5$, where  $\p {\tilde U}_{\rm eff}/\p p$ 
determines  the overall behavior  of $h$.
The resultant  $h$ behaves in the same manner as in the 
 the experiment\cite{Desnoyers}. 
Here,     the two terms in Eq.(60) compete delicately 
 depending  on  the parameter values. 
  Indeed,  if we set $A_{\rm B}=-5.0$  with 
the other parameters unchanged,  the curves 
of  $\alpha_3=0.7, 0.9,$ and 1.1 
increase with increasing $\alpha_2$ for $\alpha_2\gs 0.8$.   
We also set  $n_1\p \ln  \kappa_{\rm w}/\p n_1= -8.3$, 
   from Eqs.(80) and (81), though it is    $ -5.4$ 
in real water\cite{Millerocompress}.  
Thus, to calculate $h$, we  need  to make   very crude   
 approximations\cite{RedlichReview}.

\begin{figure}[tbp]
\begin{center}
\includegraphics[width=235pt]{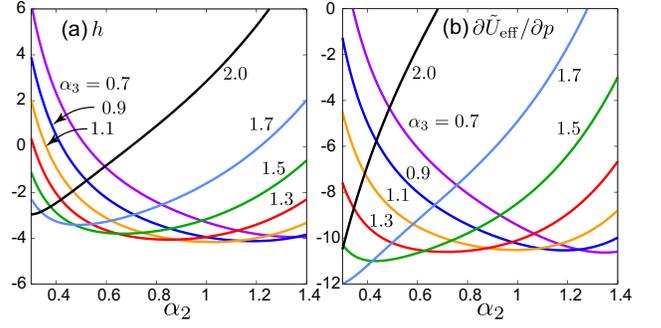}
\caption{(Color online) (a) Deviation constant  $h$ 
in Eq.(60) and (b) $\p {\tilde U}_{\rm eff}/\p p$ 
(the second term in Eq.(60)) 
vs $\alpha_2$   in units of $d_1^6$, where 
   $\alpha_3=0.7, 0.9, 1.1, 1.3, 1.5, 1.7$, 
and 2.0 from above. 
 }
\end{center}
\end{figure}

\begin{table}[tbh]
\caption{
Data  of $h$ 
for alkali halides\cite{Desnoyers}
 and Et$_4$N$^+$ halides\cite{DesnoyersT} 
in units of $\mathrm{cm}^3\mathrm{L}/\mathrm{mol}^2$ 
and in units of  $d_1^6$ in the parentheses $()$.  
 Here,  1 $\mathrm{cm}^3\mathrm{L}/\mathrm{mol}^2$ 
corresponds to $3.82d_1^6$ with $d_1=3$\AA. 
Ion volume of  Et$_4$N$^+$ is 
$v_2^*\sim 9d_1^3$ $\sim 8n_{\rm w}^{-1}$. 
\label{Tab_hv}}
\begin{ruledtabular}
\begin{tabular}[t]{c | c c c c}
& $\mathrm{F}^-$   & $\mathrm{Cl}^-$ & $\mathrm{Br}^-$ & $\mathrm{I}^-$\\
\hline
$\mathrm{Li}^+$ &1.1\ (4.2) &-0.36\ (-1.4) &-0.60\ (-2.3) & \\
$\mathrm{Na}^+$ &0.64\ (2.4) &-0.03\ (-0.11) &-0.26\ (-0.99) &-0.38\ (-1.5) \\
$\mathrm{K}^+$ &0.52\ (2.0) &0.10\ (0.38) &-0.16\ (-0.61) &-0.39\ (-1.5) \\
$\mathrm{Rb}^+$ &0.55\ (2.1) &0.17\ (0.65) &-0.26\ (-0.99) &-0.05\ (-0.19) \\
$\mathrm{Cs}^+$ &0.25\ (0.95) &0.12\ (0.46) &0.09\ (0.34) &0.11\ (0.42) \\
$\mathrm{Et}_4\mathrm{N}^+$ & & -21.0\ (-80) & -19.4\ (-74) & -6.0 (23) \\
\end{tabular}
\end{ruledtabular}
\end{table}

\section{Summary and Remarks}

In summary, we have presented a theory of electrolytes 
accounting for  the deviation of the solvent density 
$\delta n_1$ induced by those of the ions. It  has been neglected  
in the previous primitive theories.  
In Sec.III, we have then derived  the ion volume $v_i^*$ in Eq.(19) 
and the effective ion-ion interaction 
coefficients $U_{ij}^{\rm eff}$ in   Eq.(24) ($i, j=2,3)$.   
In the latter, the second bilinear term ($ - v_i^* v_j^*/\kappa_{\rm w}$) 
arises from the solvent-mediated interactions 
and can   explain   Collins'  rule\cite{Collins} in the presence of the 
electrostriction (which leads to $v_i^*<0$ for small ions). 
Namely, it yields 
 cation-anion repulsion for small-large ion pairs  
with $v_i^*v_j^*<0$ and attraction for symmetric pairs 
with $v_i^*v_j^*>0$. 
In the thermodynamic quantities, the mean interaction 
coefficient  $U_{\rm eff} = \sum_{i,j=2,3} U_{ij}^{\rm eff}/2$ appears.  										
						
We have  defined  a parameter $\chi$  
 in  the ionic fluctuation variances for $n_2$ and $n_3$ in Eq.(28) 
and  expressed  the Kirkwood-Buff integrals for $n_1$ and $n_2+n_3$ 
in terms of $\chi$ in Eqs.(36) and (37). 
We have expanded this $\chi$, 
the  mean activity coefficient $\gamma_\pm$, 
the osmotic coefficient $\varphi$, 
and the apparent partial volume $v_i^{\rm ap}$ 
in powers of $\sqrt{n_{\rm s}}$ for small average 
salt density $n_{\rm s}=\av{n_2}= \av{n_3}$.  
In these expressions the first correction are 
the DH contributions.

 We have also  confirmed  unique 
 behavior of  small-large ion pairs as predicted by Collins, 
 where  $U_{\rm eff}<0$ and  $U_{23}^{\rm eff}>0$. 
  As an extreme example, NaBPh$_4$ is  
 strongly coupled with the water density with a 
 largely negative $U_{\rm eff}$. For such a salt, 
we have  discussed a spinodal  instability 
 for  $n_{\rm s}$ exceeding   $n_{\rm s}^{\rm spi}$ 
in Eq.(68)\cite{OkamotoOnukiTernary}.

 In Sec.IV,  we have  performed  numerical analysis 
using   the Mansoori-Carnahan-Starling-Leland (MCSL) 
model\cite{MCSL},  the Lennard-Jones (LJ)  attraction,   
  and   the Born model. 
We have  calculated the ion volume $v_i^*$ 
and the excess  coefficients $V_{ij}^{\rm eff}=
 U_{ij}^{\rm eff}-  u_{ij}^{\rm ex}$ in Eq.(91)  in Fig.7,  
where  $u_{ij}^{\rm ex}$ are  the 
contribution from the  DH free energy  in Eq.(6).  
Some asymptotic expressions   have been given for them 
in Eqs.(90), (93), and (94). 
Regarding the ion-specific thermodynamic behavior,
 the mean interaction coefficient $U_{\rm
eff}=\sum_{i,j=2,3}U_{ij}^{\rm eff}/2$ is a key 
quantity (see Eqs.(56)-(62))

We have found that  the 
two steric  parts in $U_{ij}^{\rm eff}$ 
in Eq.(24) or  $V_{ij}^{\rm eff}$  in Eq.(91) 
mostly cancel, as  calculated in Appendix E. 
Due to this cancellation,  the effective interaction coefficients 
 for purely steric hardsphere systems,  $U_{hij}^{\rm eff}$  in Eq.(92), 
 are not large as in Fig.4 and become 
smaller than the other contributions for ambient water, leading to Eqs.(93) 
and (94).  Note that our hardcore  quantities,    $\nu_i^h$ and 
$U_{ij}^h$ in Eq.(74) and $1/\epsilon_{\rm in}^h$ 
in Eq.(81),  are enlarged  by the powers of $(1-\eta_1)^{-1}$ 
for large solvent volume fraction 
$\eta_1 (\sim 0.5$ for ambient water). In contrast,  in the  
primitive models\cite{Friedman,Lebowitz,Ebeling,Blum1,Blum2}, 
  the  total packing  fraction arises from  the ions only 
and  $U (=- k_BT\sum_{i,j} 
\int d{\bi r} c_{ij}^0(r)/2$)  
 is positive and  not very large (see Eq.(22)), 
so its expression (without the second term in 
Eq.(25))  was  fitted to  
  data of salts.   

 We have examined  the  Born 
part  in $V_{ij}^{\rm eff}$, which yields 
 singular interaction for small-large ion pairs. 
The remaining part consists  of the MCSL and LJ contributions 
exhibiting  rather simple behaviors in Fig.8.  
We have then presented  simple interpolation formulas  
for $V_{ij}^{\rm eff}$ in Eqs.(97)-(99). 
We have calculated $\chi^{-1}$, $w_\rho$,  $\ln \gamma_{\pm}$, 
and $\varphi$    as functions of 
 $\alpha_2$,  $\alpha_3$, and $n_{\rm s}$ in Figs.9 and 10. 
We have also examined  the deviation constant  $h$ in  
 $v_{\rm s}^{\rm ap}$ in Fig.11, which behaves 
differently from the others.


We make some  remarks. \\
(i)  Our numerical analysis is very  approximate.  
In particular, the parameter choices in Eqs.(83)-(86) 
remain still arbitrary, where  the specific 
 properties of  cations and anions are neglected.  
 Nevertheless, our theory   provides simple, overall  understanding of 
the puzzling behaviors of electrolytes. 
The results  in Fig.7 should be  commonly expected for 
 various solvents (see Appendix F). 
(ii) We should  calculate the structure factors 
of water and   ions  at finite wave numbers 
including   the DH interaction and the 
effective mutual interactions. 
(iii) It is informative to perform  molecular dynamics 
simulations for  various ion pairs, for example, 
to confirm the behaviors in Fig.7 and Eqs.(97)-(99). 
(iv) We have mentioned 
singular behaviors of   small-large ion pairs in water\cite{Collins}, 
which include  antagonistic salts\cite{Onuki2} such as NaBPh$_4$. It is  
of great interest to perform scattering experiments\cite{Collins1}  
for     salts   with small or negative $\chi^{-1}$. 
(v) In mixture  solvents such as water-alcohol, 
the solvent-mediated interaction 
is  much enhanced    due to the 
concentration fluctuations\cite{OkamotoOnukiTernary}. 
Thus, we need to study  
electrolytes of mixture  solvents. 


\begin{acknowledgments}
RO would like to thank Tomonari Sumi for informative discussions. RO acknowledges support from JSPS KAKENHI Grant (No. JP18K03562 and JP18KK0151).  KK acknowledges support from JSPS KAKENHI Grant (No. JP18KK0151 and JP20H02696). 
AO would like to thank Zhen-Gang Wang for informative correspondence.
\end{acknowledgments}
\vspace{2mm}
\noindent{\bf AIP Publishing Data Sharing Policy}\\
The data that support the findings of this study are avail- able from the corresponding author upon reasonable re- quest.

\vspace{2mm}
\noindent{\bf Appendix A: Bjerrum dipoles}\\
\setcounter{equation}{0}
\renewcommand{\theequation}{A\arabic{equation}}

Here, we  examine   how the Bjerrum dipoles  alter  our theory.
For nonvanishing  dipole density  $n_{\rm d}$, 
we change the free energy density  $f$ in Eq.(2)  to\cite{Roij}
\be
{\tilde f}=f+  k_BT n_{\rm d}      
[ \ln (n_{\rm d}\lambda_{\rm d}^3)-1  - \nu_{\rm d}+2\nu ],
\en  
where $k_BT \nu_{\rm d}$ is the free energy 
decrease due to the   association per dipole and  
 $\nu$ is  defined by Eq.(8). 
If we neglect inhomogeneous density deviations, 
we have $n_2= n_3 = n_{\rm s}-n_{\rm d}$, where $n_{\rm s}$ is 
the  added salt density (held fixed here).  In equilibrium, 
the dipole chemical potential $\mu_{\rm d}=\p {\tilde f}/\p n_{\rm d}$ 
 equals   $\mu_{\rm s}$ in Eq.(41); then, 
\be 
 {n_{\rm d}}
= K n_{\rm s}^2+\cdots, 
\quad {n_2}= {n_3}=
 n_{\rm s}-K n_{\rm s}^2+\cdots .
\en  
where  $n_{\rm s}\ll K^{-1}$  with $K$ being  the association constant,  
\be 
K= (\lambda^6/\lambda_{\rm d}^3) \exp({\nu_{\rm d}}). 
\en 
If $n_{\rm d}$ is removed,  the free energy density is lowered as  
\be 
{\tilde f}(n_1, n_2, n_3,n_{\rm d}) 
= {f}(n_{\rm w}, n_{\rm s}, n_{\rm s})  -k_BT K n_{\rm s}^2+\cdots,
\en  
where the logarithmic term $k_BT n_{\rm d}\ln (n_{\rm d}\lambda_{\rm d}^3)$ 
disappears. Thus, if we accept Bjerrum's assumption, 
$U_{23}$ in Eq.(3)  is changed to  
  $U_{23}- k_BT K$. Then,  $\ln\gamma_{\pm}$  decreases by $Kn_{\rm s}$.

\vspace{2mm}
\noindent{\bf Appendix B: Gibbs free energy of electrolytes}\\
\setcounter{equation}{0}
\renewcommand{\theequation}{B\arabic{equation}}

We  calculate  the Gibbs free energy $G$.  
 As  in Sec.IIID,  we   fix  $p$, $T$, and 
 the total solvent number $N_{\rm w}=V n_{\rm w}=V_0n_{\rm w}^0$. 
Here, without salt at pressure $p$,    the   solvent density 
is $n_{\rm w}^0$ and the volume is   $V_0$.

We   integrate   $dG/dN_{\rm s}=\mu_{\rm s}$ 
 with respect to $N_{\rm s}$ using  Eq.(41), 
where we set  $n_{\rm s}= N_{\rm s}/V\cong (N_{\rm s}/V_0)
( 1- {\bar v}_{{\rm s}}^0N_{\rm s}/V_0)$ in $\ln(n_{\rm s}\lambda^3)$. 
Up to order $n_{\rm s}^2$, we  obtain\cite{RBook,Pitzer}   
\bea 
&&\hspace{-4mm}
G= N_{\rm w}\mu_{\rm w}^0(n_{\rm w}^0)  
 +2 k_BTN_{\rm s} [\ln (\lambda^3N_{\rm s}/V_0) -1+ 
\nu(n_{\rm w}^0)]\nonumber\\
&& \hspace{3mm} 
+V_0( -k_BT \kappa^3/12 + {\tilde U }_{\rm eff} n_{\rm s}^2),   
\ena
where    $\mu_{\rm w}^0(n_{\rm w}^0) $ 
is the chemical potential of pure solvent at the 
density $n_{\rm w}^0 $   and ${\tilde U }_{\rm eff}$ is given in Eq.(57).  

Since  $n_{\rm w}^0$ is determined by 
$p$ and $T$, we can treat $G$ in Eq.(B1) as a 
function of $N_{\rm w}$, $N_{\rm s}$, 
$p$, and $T$.  Then,  
\be
V= (\p G/\p p)_{N_{\rm w}, N_{\rm s},  T}=
 V_0+ v_{\rm s}^{\rm ap} N_{\rm s},    
\en   
from which we can calculate the apparent partial volume 
$v_{\rm s}^{\rm ap}$ to derive   Eqs.(59) and (60) up 
order $n_{\rm s}^2$.  The partial volume  
${\bar v}_{\rm s}$ in Eq.(46)  
can be related to   $v_{\rm s}^{\rm ap}$ as  
\be 
{\bar v}_{\rm s}= [v_{\rm s}^p+ n_{\rm s} (\p v_{\rm s}^p/\p n_{\rm s})]
/[1+ n_{\rm s}^2 (\p v_{\rm s}^p/\p n_{\rm s})],
 \en 
where   $\p v_{\rm s}^p/\p n_{\rm s}$ is the derivative at fixed $p$ and $T$. 
On the other hand, from   $G= N_{\rm w}\mu_{\rm w}+ N_{\rm s}\mu_{\rm s}$, 
the osmotic coefficient $\varphi$ in  Eq.(61) is expressed as 
\be
\varphi= [  N_{\rm w} \mu_{\rm w}^0(n_{\rm w}^0) 
 + N_{\rm s}\mu_{\rm s}-G]/(2k_BT  N_{\rm s}), 
\en  
leading to  Eq.(62)  with the aid of   Eqs.(41) and (B1).

\vspace{2mm}
\noindent{\bf Appendix C:Derivation of Eqs.(50) and (59) }\\
\setcounter{equation}{0}
\renewcommand{\theequation}{C\arabic{equation}}

We rewrite  Eqs.(49) and (58) as  
\be 
dn_{\rm w}/dn_{\rm s}= b_1(n_{\rm w}) +b_2(n_{\rm w})n_{\rm s}^{1/2}+ 
b_3(n_{\rm w})n_{\rm s} +\cdots ,
\en 
where $b_1$, $b_2$, and $b_3$ are 
functions of $n_{\rm w}$.  Up to order $n_{\rm s}^2$, Eq.(C1) 
yields   the deviation 
$\delta n_{\rm w}= n_{\rm w}(n_{\rm s}) - n_{\rm w}^0$ 
 as  
\bea 
&&\hspace{-8mm} 
\delta n_{\rm w}=  b_1(n_{\rm w})n_{\rm s}
 + \frac{2}{3} b_2(n_{\rm w})n_{\rm s}^{3/2}+ \frac{1}{2} 
c_3(n_{\rm w})n_{\rm s}^2 +\cdots \nonumber\\
&&\hspace{-1mm}=b_1(n_{\rm w}^0)n_{\rm s}
 +\frac{2}{3} b_2(n_{\rm w}^0)n_{\rm s}^{3/2}+ \frac{1}{2} 
{\tilde c}_3(n_{\rm w}^0)n_{\rm s}^2 +\cdots, 
\ena 
where the second line is written in terms of $n_{\rm w}^0$ 
as in Eqs.(50) and (59).  Thus, $b_1(n_{\rm w})
\cong  b_1(n_{\rm w}^0)+ b_1' \delta n_{\rm w}$, 
where $b_1'= d b_1/d n_{\rm w}$. 
We differentiate the first line of Eq.(C2) with respect to $n_{\rm s}$ 
to  find  
\be 
  c_3= b_3- b_1 b_1', \quad 
{\tilde c}_3= b_3+ b_1 b_1'.   
\en 
The expression for ${\tilde c}_3$ leads to 
 Eqs.(50) and (59).

\vspace{2mm}
\noindent{\bf Appendix D:  Extended expressions 
of $\chi^{-1}$ and $v_{\rm s}^{\rm ap}$ }\\
\setcounter{equation}{0}
\renewcommand{\theequation}{D\arabic{equation}}

We rewrite $\chi^{-1}$ in Eq.(32)  
and $v_{\rm s}^{\rm ap}$ in Eq.(59) as 
\bea 
&&\hspace{-2mm} \chi^{-1}= 2-  \frac{1}{4}
\ell_B \kappa \sum_{i=2,3} \frac{1}{(1+ a_i \kappa)^{2}} 
 + \frac{2n_{\rm s}}{k_BT} V_{\rm eff} ,\\
&&\hspace{-3mm} 
{v}_{\rm s}^{\rm ap} = {\bar v}_{{\rm s}}^0(n_{\rm w}^0) +
{\epsilon_{\rm in} {\ell}_B  \kappa}  
  \sum_{i=2,3} \Big[ 
\frac{  \epsilon'/2\epsilon}{1+ a_i \kappa} 
-\frac{ \sigma(a_i \kappa)}{6n_{\rm w}}  
 \Big] \nonumber\\
&&\hspace{4mm} + [ \kappa_{\rm w} {\tilde V}_{\rm eff}
+ \p   {\tilde V}_{\rm eff}/\p p] n_{\rm s} .
\ena  
We define   $V_{\rm eff}$ in Eq.(91) 
 and  ${\tilde V}_{\rm eff}$ in  Eq.(57). 
In Eq.(D2),   the first term is ${\bar v}_{{\rm s}}^0$   
at the initial density $n_{\rm w}^0$  and  
  $\sigma(x)$ is defined below Eq.(65). 
These expressions tend to Eqs.(32) and (59) as $a_i \kappa \to 0$.

\vspace{2mm}
\noindent{\bf Appendix E: MCSL  model of hardsphere fluids}\\
\setcounter{equation}{0}
\renewcommand{\theequation}{E\arabic{equation}}

Here, we     summarize  the MCSL   model of hardsphere fluid 
 mixtures of $m$ components\cite{MCSL}, where $m=3$ in this paper. 
Setting $n=\sum_i n_i$, $\eta_i={\pi} n_i d_i^3/6$, 
   $\eta = \sum_j \eta_j$, and $u=\eta/(1-\eta)$, we write  
$f_h(n_1,n_2,n_3) $    in Eq.(73) as\footnote{In the original 
paper\cite{MCSL},  another quantity $y_2$ also appears. In Eq.(E1), 
it is removed from the relation $y_2 = 1-y_1-y_3$.}$^,$\cite{Wa}      
\be
\frac{f_h}{k_B Tn} = 4u+u^2 -3y_1 u +(y_3-1)[u+{u^2}+ \ln (1-\eta)].  
\en 
Setting     $  \sigma_\ell  = \sum_i \pi d_i^\ell  n_i/6$, we 
  define  $y_1$ and $y_3$  as 
\be
y_1 = 1- 6\sigma_1\sigma_2/(\pi n\eta),~~
y_3 ={6 \sigma_2^3}/({{\pi \eta^2n}}), 
\en 
where  $y_1 \to 0$ and $y_3\to 1$ in the one-component limit.

From Eq.(E1) we obtain  the MCSL chemical 
potentials $\p f_h/\p n_i$. In the dilute case, 
  $\nu_i^h$ in Eq.(74) are written as      
\begin{align}
 \nu_i^{\rm h}= &(3\alpha_i +6\alpha_i^2-\alpha_i^3 )u_1 + 
(3\alpha_i^2+4\alpha_i^3) u_1^2
 \nonumber\\
&\hspace{-3mm}+2\alpha_i^3u_1^3+ (3\alpha_i^2-2 \alpha_i^3-1)\ln (1- \eta_1),
\end{align}
where  $u_1=\eta_1/(1-\eta_1)$. The right hand side  steeply 
grows with increasing $\eta_1$ (see Fig.3 in our previous 
 paper\cite{OkamotoOnukiDFT}). 
 The MCSL ion volume $v_i^h= \epsilon_{\rm in} {d}\nu_i^h/{dn_1}$  is given by 
\be
 { v_i^h}=  (w_{\rm a}+1 -\epsilon_{\rm in}u_1^2)  \alpha_i^3v_1 
 +\epsilon_{\rm in} \psi_i v_1,
\en 
where   $v_1=\pi d_1^3/6$ and $w_{\rm a}$ is defined in Eq.(82). 
Setting $x_1=1/(1-\eta_1)$, we define $\psi_i$ by  
\be 
\psi_i= {6\alpha_i^2}x_1^3  + 
 {3\alpha_i}x_1^2 + ({1-3\alpha_i^2})x_1.  
\en   
For    $\alpha_i=1$,  we simply find 
$
v_i^h= n_1^{-1}(1+w_{\rm a}- \epsilon_{\rm in}). 
$ 
For considerably  large $\alpha_i$ (say,  
$\alpha_i\sim 1.4$), the second term in Eq.(E4) 
is of order $v_1$, but it  
is   considerably  cancelled by negative $v_i^{\rm LJ}$  
(see  Fig.3). We thus find  Eq.(89).  

From Eqs.(E1) and (74) we  express  $U_{ij}^h$   as   
\bea 
&& \hspace{-5mm} 
{U_{ij}^h}/{v_1 k_BT} =
 \alpha_i^3\alpha_j^3 \Phi_1+\alpha_{ij}^2(\alpha_{ij} 
+3\alpha_i\alpha_ju_1 )/(1-\eta_1) 
\nonumber\\
&&\hspace{-6mm} +
6\alpha_i^2\alpha_j^2 \Big[(\alpha_{ij}-1) \Big(u_1^3+2u_1^2+\zeta_1 \Big)
 +u_1/(1-\eta_1)^2 \Big],
\ena 
where $\alpha_{ij}=\alpha_{i}+\alpha_{j}$ and $\zeta_1
=-1-\eta_1^{-1} \ln (1- \eta_1)$. 
Using $\psi_i$  
we   also express   $U_{hij}^{\rm eff}$ in Eq.(92) as 
\bea  
&&\hspace{-5mm}
{U_{h ij}^{\rm eff} }/{v_1 k_BT}
=\alpha_i^3\alpha_j^3 \Phi_2 
+  3\alpha_i^2\alpha_j^2 (\alpha_{ij}-1) (2\zeta_1-{u_1})
\nonumber\\
&& + 3\alpha_i^2\alpha_j^2(3-\eta_1)u_1/(1-\eta_1) 
  + \alpha_{ij}^3+ 3\alpha_{ij}
\alpha_i\alpha_j u_1 \nonumber\\
&&+\epsilon_{\rm in}^h \eta_1  [(\psi_i\alpha_j^3+ \psi_j\alpha_i^3)u_1^2 
 - \psi_i\psi_j]   .
\ena 
As the coefficients of $\alpha_i^3\alpha_j^3$, 
we define $\Phi_1$ and $\Phi_2$ as 
\bea 
\hspace{-5mm} \Phi_1 &=&\Phi_2 + \eta_1
( 1- u_1^2\epsilon_{\rm in}^h )^2/\epsilon_{\rm in}^h, \\  
\hspace{-5mm} \Phi_2&=&{2\eta_1}/{(1-\eta_1)}+\eta_1 -6\zeta_1 
- \epsilon_{\rm in}^h \eta_1  u_1^4,  
\ena 
where  $\Phi$ is large ($\gg 1$) but   $\Phi_2$ is  small ($\ll 1$) 
for $\eta_1\sim 0.5$.    In fact, 
 $\Phi_2 \cong \eta_1^3/2$ for $\eta_1\ll 1$ 
 and  $\Phi_2 \sim 0.1$ for $\eta_1\sim 0.5$. 
Thus,  the first term in Eq.(E7) 
is negligible for not very large $\alpha_i\alpha_j$.
 For small  $\alpha_i$ and $\alpha_j$ ($\ll 1)$, 
we have ${U_{h ij}^{\rm eff} }/v_1k_BT 
\cong -\epsilon_{\rm in} \eta_1/(1-\eta_1)^2$.
We plot ${U_{h ij}^{\rm eff} }$ in Fig.4.

Now, we rewrite  $V_{ij}^{\rm eff}$ in Eq.(91)  as  
\be 
V_{ij}^{\rm eff}= {U_{h ij}^{\rm eff} }-w_{ij} -(v_i^*v_j^*- v_i^h v_j^h 
\epsilon_{\rm in}^h/\epsilon_{\rm in})/\kappa_{\rm w}, 
\en 
where the MCSL contribution is subtracted in the third   term.  
Here,  the third term  dominates over the first  
with significant attractive and hydration interactions. 
The above  expression 
leads to Eqs.(93) and (94).

\vspace{2mm}
\noindent{\bf Appendix F: Marginal ion-size-dependence }\\
\setcounter{equation}{0}
\renewcommand{\theequation}{F\arabic{equation}}

\begin{figure}[th]
\begin{center}
\includegraphics[width=215pt]{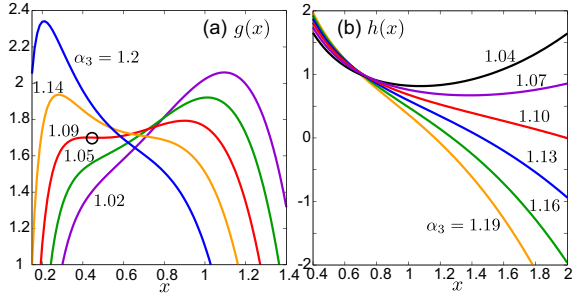}
\caption{(Color online) (a) 
$g(x)= V_{\rm eff}/(k_BTd_1^3B\alpha_3^2)$ in Eq.(F1) 
vs  $x=\alpha_2/\alpha_3$ at fixed 
$\alpha_3$. Line of $\alpha_3=1.09$ 
has an inflection point ($\circ$).  
 (b) $h(x)= V^{\rm eff}_{23}/(k_BTd_1^3 
B\alpha_3^2)$ in Eq.(F6). These functions are 
nearly flat for $\alpha_3\cong 1.1$. 
 }
 \label{scalingfunc}
\end{center}
\end{figure}

We first  show the existence of  an  inflection point 
in $V_{\rm eff}$ vs $\alpha_2$,  where $\p V_{\rm eff}/\p \alpha_2= 
\p^2 V_{\rm eff}/\p\alpha_2^2=0$ at a certain  $\alpha_3$. 
Using   Eq.(97), we  define $g(x)= V_{\rm eff}/(k_BTd_1^3B\alpha_3^2)$. 
As a function of  $x=\alpha_2/\alpha_3$ at fixed $\alpha_3$, 
$g(x)$ is  expressed as 
\be 
g(x) =(1+\frac{1}{x})\Big[(1+x^3)  -\frac{ C'}{2}(1+\frac{1}{x})\Big]- 
\frac{A'}{2} (1+x^{3})^2,
\en
where   $C'= C/B\alpha_3^4=D_{\rm B}/(D_{\rm L}\alpha_3^4)$ 
and  $A'=  A\alpha_3^4/B$ with  
 $A=5.0$, $B=7.0$, and $C=2.8$ (see  below Eq.(97)), so  
we fix $A'C'=  0.29$. We require 
$dg/dx=d^2g/dx^2=0$  at the inflection point to obtain 
\bea 
&&\hspace{-5mm} 3x^4-x^3+ x^2-x+ C'= 3A' x^5( x^2-x+1),\\
&&\hspace{-5mm} 12x^3-3x^2+ 2x-1 = 3A' x^4( 7x^2-6x+5),
\ena
The {\it critical} values of $x$ and $C'$ are 
$x_c= 0.45$ and $ C_c'= 0.25$, respectively. The critical value of 
 $\alpha_3$   is given by 
\be 
\alpha_{3c}=(D_{\rm B}/D_{\rm L}C_c')^{1/4}=1.09 .
\en 
which is  close to 1 owing to the small exponent $1/4$. 
However, $x_c$ is considerably smaller than 1, so 
the right hand sides of Eqs.(F2) and (F3) are 
negligible near the inflection point. 
For small  $\alpha_3-\alpha_{3c}$ 
and $x-x_c$, we find  
\be 
g(x) \cong 7(x-x_c)^3-16(\alpha_3-\alpha_{3c}) (x-x_c)+ 1.7 .
\en 
Thus, the slope of $g(x)$ vs $x$ changes its sign abruptly 
for $\alpha_3\cong  \alpha_{3c}$ as in  Fig.7(a), which 
  is analogous to the isothermal 
pressure-density relation in the van der Waals equation of state.

\begin{table}[tbph]
\caption{Values of $\epsilon$, $\ell_B$ ($\rm\AA$), 
$(\p \ln \epsilon/\p p)_T$ (GPa$^{-1}$), 
$d_1$ ($\rm\AA$), 
$\beta_1$, and $\alpha_{3{\rm c}}$ for six solvents at $T=300$ K.
\label{Tab_kec}}
\begin{ruledtabular}
\begin{tabular}[t]{c | c c c c c c}
& $\epsilon$ & $\ell_B$    & $\p \ln \epsilon/\p p$  & $d_1$ & $\beta_1$ 
& $\alpha_{3{\rm c}}$ \\
\hline
water &80& 7  & 0.47 & 3  & 0.91 & 1.09\\
formamide &111& 5  & 0.45 & 3.9  & 0.62 & 0.74\\
methanol& 33& 17  & 1.2 & 3.9  & 1.09& 1.31\\
ethanol &25& 22  & 1.2 & 4.4  & 1.03 & 1.24\\
acetonitrile &37& 15  & 1.1 & 4.3  & 0.96 & 1.15\\
acetone &21& 27  & 1.6 & 4.8  & 1.07 & 1.28\\
\end{tabular}
\end{ruledtabular}
\end{table}

Second,  we  consider the normalized cation-anion interaction coefficient 
  $h(x)= V_{23}^{\rm eff}/(k_BTd_1^3 
B\alpha_3^2)$. From  Eq.(99), $h(x)$ depends on  
$x=\alpha_2/\alpha_3$  as 
\be 
h(x) = (1-A') x^3+ (1-C') /x,
\en 
which  has no inflection point. 
However,  if $A'\cong 1$ or $\alpha_3\cong (B/A)^{1/4}\sim 1.1$, 
$h(x)$ is nearly flat, say, 
in the range $[0.6,1.2]$  as in Fig.~\ref{scalingfunc}(b).   
For example,  we have 
$A'= 1.05$ and 0.94 for  $\alpha_3=1.10$ and 1.07, 
respectively. 
This behavior  can be seen in Fig.7(b).

Third,  we discuss the marginal size-dependence of $V_{\rm eff}$ 
for nonaqueous solvents. From Eq.(F4) and the sentences below Eq.(90), 
we have  
$\alpha_{3{\rm c}}= \beta_1 \beta_2$ with 
\begin{align}
\beta_1=[4\ell_B\epsilon_{\rm in}\epsilon'/\epsilon ]^{1/4}/d_1, 
\quad  \beta_{2}=(d_i/2R_i D_{\rm L})^{1/4},
\end{align}
where  we set $C_c'=1/4$. We also set $\beta_2= 1.2$ as in the case 
of water, while   $\beta_1$ depends on the solvent species. 
  For nonaqueous solvents, we assume  $d_1= n_1^{-1/3}$ 
and use published experimental data  at $T\sim 300$K and $p\sim 1$ 
atm\cite{Craig,Hefter}.
We then obtain Table V, where $\alpha_{3{\rm c}}\sim 1$ 
 for all the solvents (again largely due to the  exponent $1/4$).

\bibliography{bib}

%
%
\end{document}